\documentclass{aa}  

\usepackage[backref,breaklinks,colorlinks,linkcolor=blue,urlcolor=blue,citecolor=blue]{hyperref}
\usepackage[all]{hypcap}
\usepackage[varg]{txfonts}

\usepackage{graphicx}
\usepackage{bm}
\usepackage{hyperref}
\newcommand{\msun}{\,\mathrm{M}_{\sun}}
\newcommand{\mpc}{\,\mathrm{Mpc}\,h^{-1}}

\begin{document}

\title{He abundances in disc galaxies -- I. Predictions from cosmological chemodynamical simulations}

\author{Fiorenzo Vincenzo\inst{1}, Andrea Miglio\inst{1}, Chiaki Kobayashi\inst{2}, J. Ted Mackereth\inst{1}, Josefina Montalban\inst{1}} 
\institute{School of Physics and Astronomy, University of Birmingham, Edgbaston,  B15 2TT, UK \\
\email{f.vincenzo@bham.ac.uk} \and 
Centre for Astrophysics Research, University of Hertfordshire, College Lane, Hatfield, AL10 9AB, UK 
}




\abstract{We investigate how the stellar and gas-phase He abundances evolve as functions of time within simulated star-forming disc galaxies with different star formation histories. We make use of a cosmological chemodynamical simulation for galaxy formation and evolution, which includes star formation, as well as energy and chemical enrichment feedback from asymptotic giant branch stars, core-collapse supernovae, and Type Ia supernovae. The predicted relations between the He mass fraction, $Y$, and the metallicity, $Z$, in the interstellar medium of our simulated disc galaxies depend on the past galaxy star formation history. In particular, $dY/dZ$ is not constant and evolves as a function of time, depending on the specific chemical element that we choose to trace $Z$; in particular, $dY/dX_{\text{O}}$ and $dY/dX_{\text{C}}$ increase as functions of time, whereas $dY/dX_{\text{N}}$ decreases. In the gas-phase, we find negative radial gradients of $Y$, due to the inside-out growth of our simulated galaxy discs as a function of time; this gives rise to longer chemical enrichment time scales in the outer galaxy regions, where we find lower average values for $Y$ and $Z$. Finally, by means of chemical evolution models, in the galactic bulge and inner disc, we predict steeper $Y$ versus \textit{age} relations at high $Z$ than in the outer galaxy regions. We conclude that, for calibrating the assumed $Y$-$Z$ relation in stellar models, C, N, and C+N are better proxies for the metallicity than O, because they show steeper and less scattered relations. 
} 


\titlerunning{He abundances in chemodynamical simulations}
\authorrunning{F. Vincenzo et al.}

\maketitle

\section{Introduction} \label{sec:intro}

In order to study the star formation history (SFH) of our Galaxy, undestanding how He abundances are distributed in the stars and interstellar medium (ISM) is fundamental for a precise estimate of stellar ages for different metallicity and star-formation enviroments \citep{iben1968,chiosi1982,fields1996,chiappini2002,jimenez2003,romano2007}.
Stellar ages in our Galaxy are typically inferred either by fitting the observed colour-magnitude diagram (CMD) with a set of stellar isochrones (e.g., \citealt{bensby2014}), or by adding asteroseismic constraints \citep{casagrande2016,silvaaguirre2016,miglio2017,silvaaguirre2018}. In both cases the resulting age distributions rely on the assumptions and underlying physics of stellar models \citep{casagrande2007,lebreton2014}.
One of the most important assumptions of stellar models is given by the initial He content of the stars, and stellar models are usually calibrated by assuming a linear scaling relation between the He mass fraction, $Y_{\star}=M_{\text{He}}/M_{\star}$, and the stellar metallicity, $Z_{\star}$ (which may be obtained from absorption lines in stellar spectra) (e.g., \citealt{pagel1998}). 

Historically, helium abundances in the stars of our Galaxy could be directly measured only in the photospheres of O- and B-type stars  \citep{struve1928,shipman1970,morel2006,nieva2012}, because in the later  spectral types there are no strong He absorption features for accurate spectroscopic analysis. Helium abundances in the ISM of galaxies can be directly measured within Galactic and extragalactic HII regions, by making use of optical He recombination lines (e.g., HeI $\lambda4471$, $\lambda5875$; see, for example, \citealt{peimbert2017}); in metal-poor HII regions, these observations have been instrumental to determine (via extrapolation towards zero metallicity) an independent estimate of the primordial He abundance to compare with the predictions of the Big Bang nucleosynthesis (e.g., \citealt{peimbert1976,peimbert2002,luridiana2003,olive2004,izotov2007,valerdi2019}). Another viable mean of measuring He abundances in the HII regions is by using the He radio recombination lines in the $8$-$10\;\text{GHz}$ frequency interval \citep{balser2006}. Finally, He abundance measurements can also be obtained in planetary nebulae (e.g., \citealt{pottasch2010,stanghellini2006}) and in absorptions systems along the line-of-sight to quasars at high redshift \citep{cooke2018}.

It is interesting to note that \citet{aver2015} showed that the HeI $\lambda10830$ emission line in the near-infrared -- when combined with high-quality optical data -- can be used to greatly reduce the uncertainties in the abundance analysis; in fact, HeI $\lambda10830$ is strongly sensitive to the HII region electron density \citep{izotov2014}, and this can be used to break the density-temperature degeneracy in the abundance analysis.

\begin{figure}[t!]
\centering
\includegraphics[width=7cm]{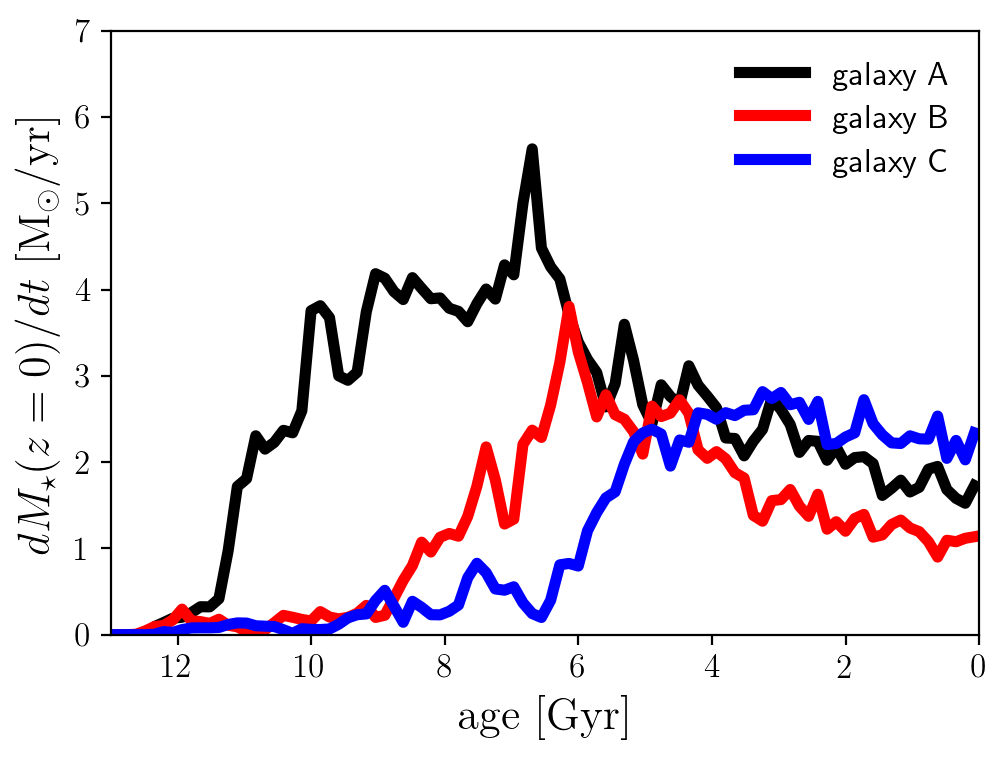} 
\caption{ The predicted age distribution of the stellar populations in our three reference galaxies. The curves are such that the area below each of them corresponds to the galaxy stellar mass at redshift $z=0$. 
}
\label{fig0}
\end{figure}

 \begin{figure}[t!]
\centering
\includegraphics[width=8cm]{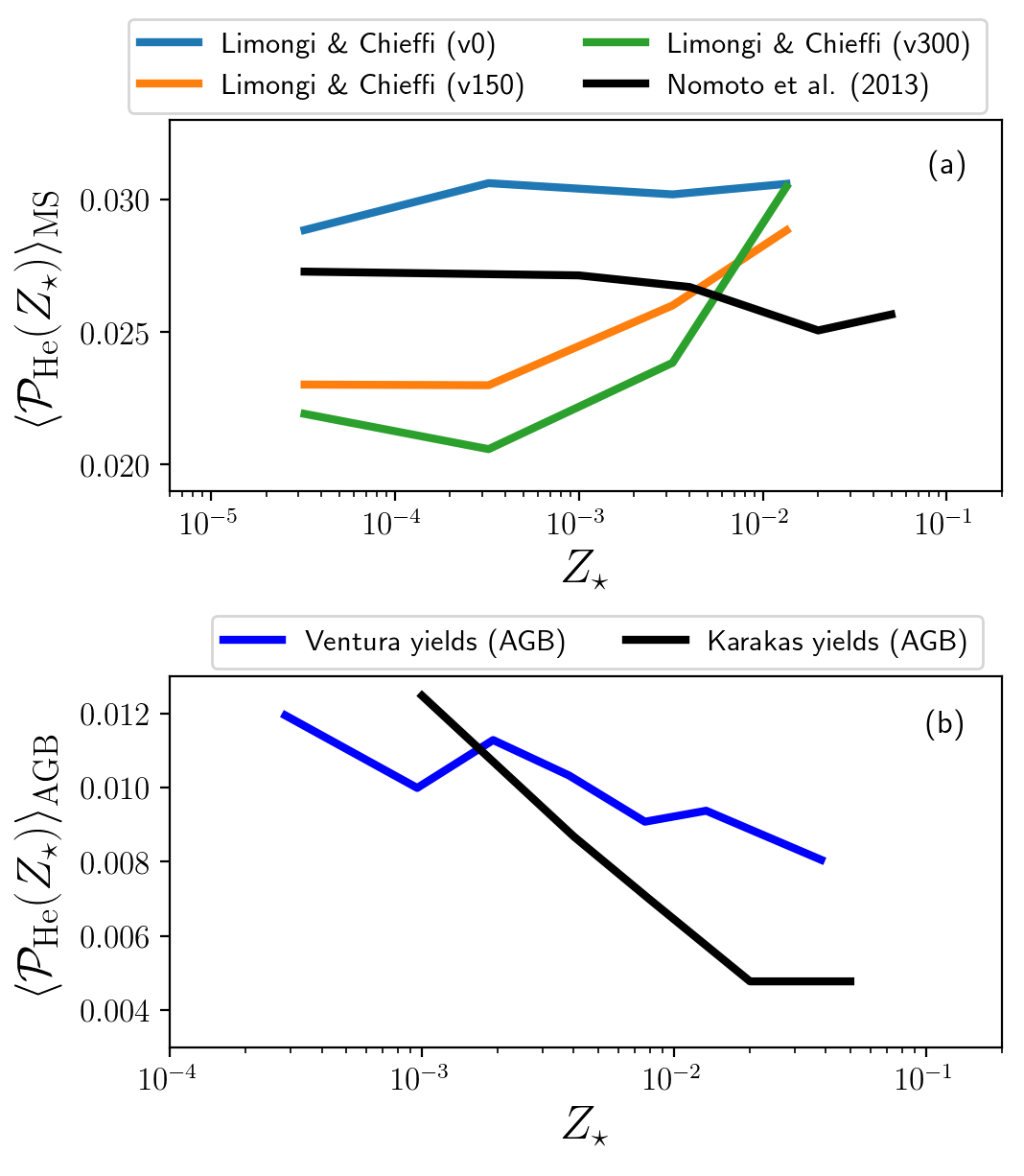} 
\caption{
\textit{(a)} The average IMF-weighted stellar yields of He from massive stars as functions of the initial stellar metallicity, as predicted by the stellar models of \citet[black solid line]{nomoto2013} and \citet{limongi2018} for different stellar rotational velocities; in particular, the blue, green, and orange lines correspond to the \citet{limongi2018} stellar models with $v_\text{rot} = 0$, $150$, and $300\,\text{km}\,\text{s}^{-1}$, respectively. \textit{(b)} The average IMF-weighted stellar yields of He from AGB stars as functions of the initial stellar metallicity, as predicted by the stellar models of \citet[black solid line]{karakas2010} and \citet[blue solid line]{ventura2013}.
}
\label{fig:heliumcontribution}
\end{figure}

 \begin{figure}[t!]
\centering
\includegraphics[width=9cm]{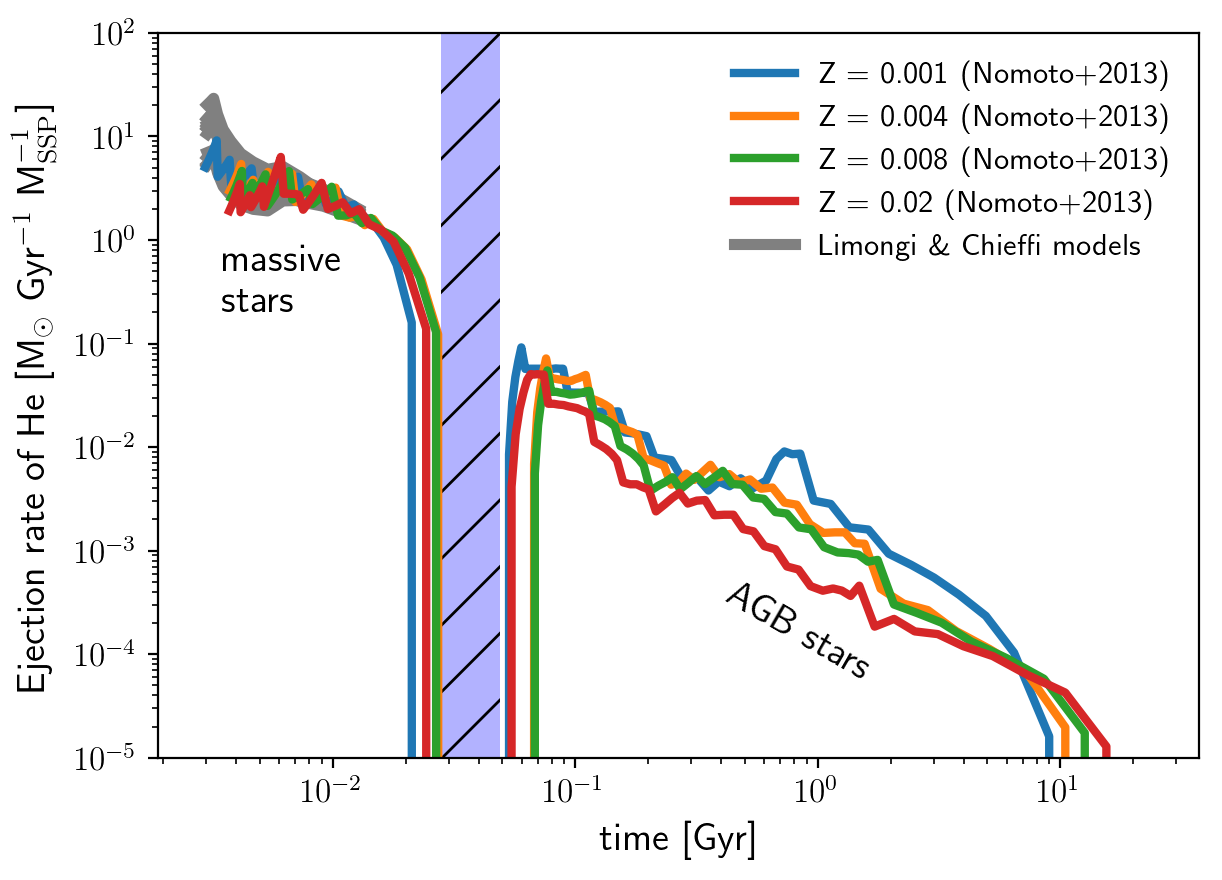} 
\caption{The predicted ejection rate of He from stellar populations of different metallicities (different colours) by assuming the stellar yields of \citet{nomoto2013}. The gray shaded area corresponds to our predictions when assuming the \citet{limongi2018} set of stellar yields of rotating massive stars with different metallicities ($\text{[Fe/H]}=0$, $-1$, $-2$, and $-3\,\text{dex}$) and stellar rotational velocities ($v_{\text{rot}}=0$, $150$, and $300\,\text{km}\,\text{s}^{-1}$). The blue shaded horizontal region corresponds to the chemical enrichment of super-AGB stars which are not included in our galactic chemical evolution model. 
}
\label{fig:heliunrate}
\end{figure}

Putting together the results of all studies, one finds a large variations for $dY/dZ$, ranging in the interval $[1.5,2.5]$\footnote{We address the readers to the following website hosted by the University of Rochester for a referenced list of measured $dY/dZ$ from different sources in the literature, put together by Eric Mamajek: \url{http://www.pas.rochester.edu/~emamajek/memo_dydz.html}.}. 
This makes the best calibration for $Y=Y(Z,t)$ to assume in stellar models highly uncertain, from both an observational and theoretical point of view \citep{casagrande2007,portinari2010,gennaro2010}; this results -- in turn -- in a large uncertainty in the final estimate of stellar ages. In particular, the impact of assumed $dY/dZ$ may give rise to relative variations in the measurement of stellar ages from asteroseismic analysis as high as $\sim40$ per cent (\citealt{lebreton2014}, Miglio et al., in prep.).
Moreover, \citet{nataf2012} pointed out that there may be critical issues in determining the ages of Galactic bulge stars if $dY/dZ \ne \text{const}$ (see also \citealt{renzini2018}). 

It is therefore important to investigate how the He content in the ISM and in the stars of galaxies depends on the metallicity and SFH. In this paper, we present the first attempt to study how He is produced and then released by ageing stellar populations in galaxies, by making use of a state-of-the-art cosmological chemodynamical simulation \citep{vincenzo2018a,vincenzo2018b}.

Our paper is structured as follows. In Section \ref{sec:model}, we summarise the main characteristics of our cosmological hydrodynamical simulation. In Section \ref{sec:helium}, we describe how He is deposited by ageing stellar populations of different metallicities in our simulation, trying to quantify the uncertainty due to different stellar yield assumptions by means of a one-zone chemical evolution model. In Section \ref{sec:results}, we present the results of our study on He in the ISM and in the stars of our simulated galaxies. Finally, in Section \ref{sec:conclusions}, we draw our conclusions.


\section{The assumed cosmological simulation} \label{sec:model}

In this paper we make use of the same cosmological hydrodynamical simulation code as described in detail in \citet{kobayashi2007,vincenzo2018a,vincenzo2018b}, which is based on the \textsc{Gadget-3} code \citep{springel2005}. 
Our simulation code includes star formation activity with mass- and metallicity-dependent chemical and thermal energetic feedback from \textit{(i)} stellar winds of dying asymptotic giant branch (AGB) and massive stars of all masses and metallicities, \textit{(ii)} Type Ia Supernovae (SNe), \textit{(iii)} Type II SNe and hypernovae (HNe).

We assume the same nucleosynthesis prescriptions as in \citet{kobayashi2011b}, modified to include failed SNe for masses $m \ge 25\,\text{M}_\sun$ and metallicities $Z \ge 0.02$ of the progenitor massive stars (see also \citealt{smartt2009}, \citealt{muller2016}, \citealt{beasor2018}, \citealt{prantzos2018}, \citealt{limongi2018}, \citealt{sukhbold2019}, and Kobayashi et al., in prep.). 
Finally, the initial mass function (IMF) is that of \citet[similar to \citealt{chabrier2003}]{kroupa2008}, defined between $0.01$ and $120\,\text{M}_{\sun}$.

 \begin{figure*}[t!]
\centering
\includegraphics[width=18cm]{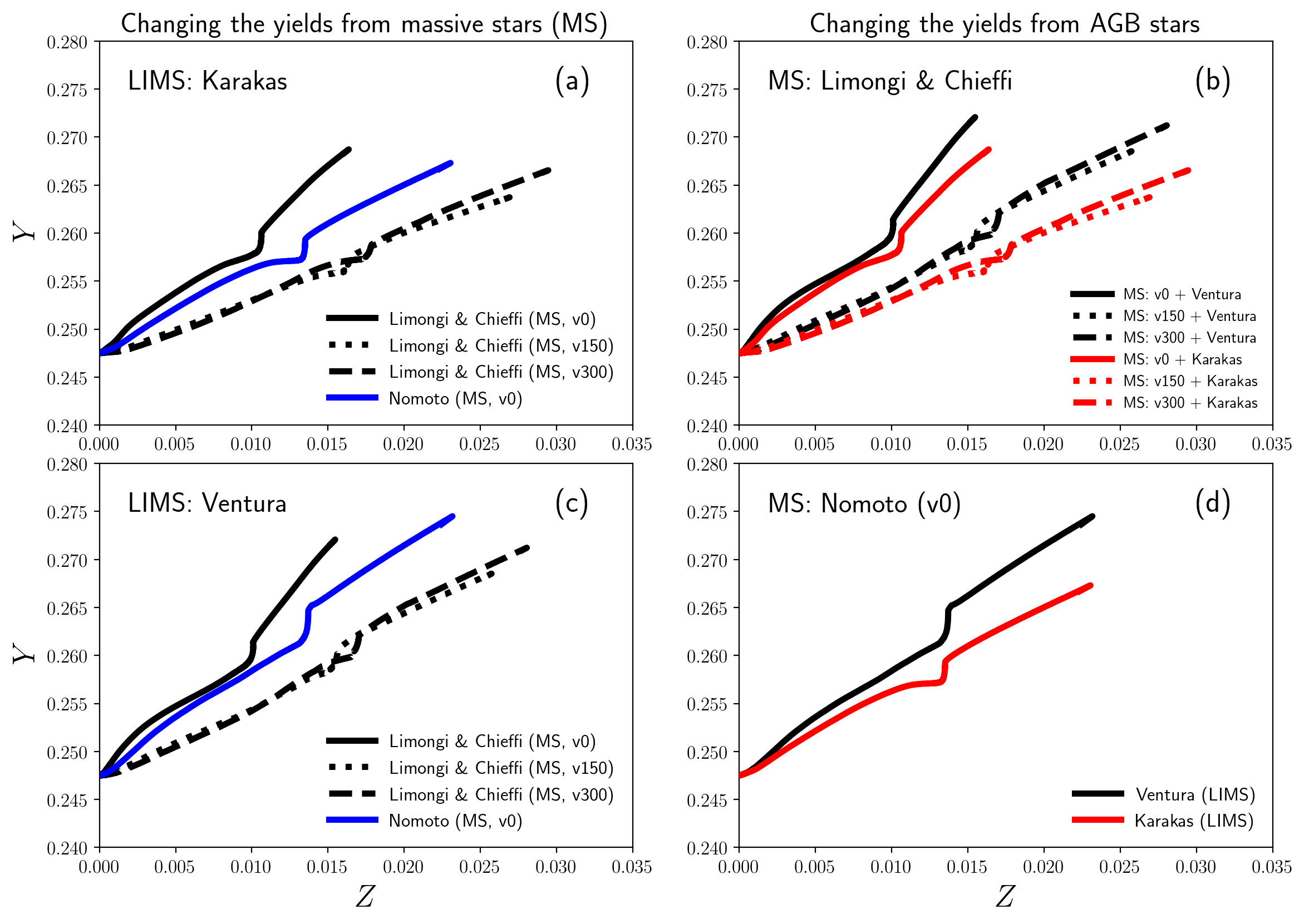} 
\caption{In this figure, we explore the effect of different nucleosynthesis yields for massive stars (panels a and c) and AGB stars (panels b and d) on the predicted $Y$ versus $Z$ relation from a one-zone galaxy chemical evolution model with star formation efficiency $\text{SFE}=2\,\text{Gyr}^{-1}$, infall time scale $\tau=4\,\text{Gyr}$, and no galactic winds. For massive stars we explore the stellar yields of \citet{nomoto2013}, which are adopted in our cosmological simulations, and the stellar yields of \citet{limongi2018} for different stellar rotational velocities ($v0$, $v150$ and $v300$ represent rotational velocity $v=0$, $150$, and $300\,\text{km}\,\text{s}^{-1}$, respectively). For AGB stars, we explore the stellar yields of \citet{karakas2010}, adopted in our cosmological simulations, and those of \citet{ventura2013} (e.g., see \citealt{vincenzo2016}). 
}
\label{fig:tracks}
\end{figure*}

In our simulation, we evolve a cubic volume of the $\Lambda$-cold dark matter ($\Lambda$CDM) Universe, with the cosmological parameters being given by the nine-year Wilkinson Microwave Anisotropy Probe \citep{hinshaw2013}. 
We assume periodic boundary conditions and 
a box side $\ell_{\mathrm{U}}=10\mpc$, in comoving units.  
We have a total number of gas and DM particles which is  $N_{\mathrm{DM}}=N_{\mathrm{gas}}=128^3$, with the following mass 
resolutions: $M_{\mathrm{DM}} \approx3.097\times10^{7}\,h^{-1}\msun$ and $M_{\mathrm{gas}}=6.09\times10^{6}\,h^{-1}\msun$. Finally, the gravitational 
softening length is $\epsilon_{\mathrm{gas}}\approx0.84\,h^{-1}\;\mathrm{kpc}$, in comoving units.
Our simulations gives a good agreement with the observed cosmic SFR, mass-metallicity relations, and the N/O--O/H relation \citep{vincenzo2018b}.

\begin{figure*}[t]
\centering
\includegraphics[width=18cm]{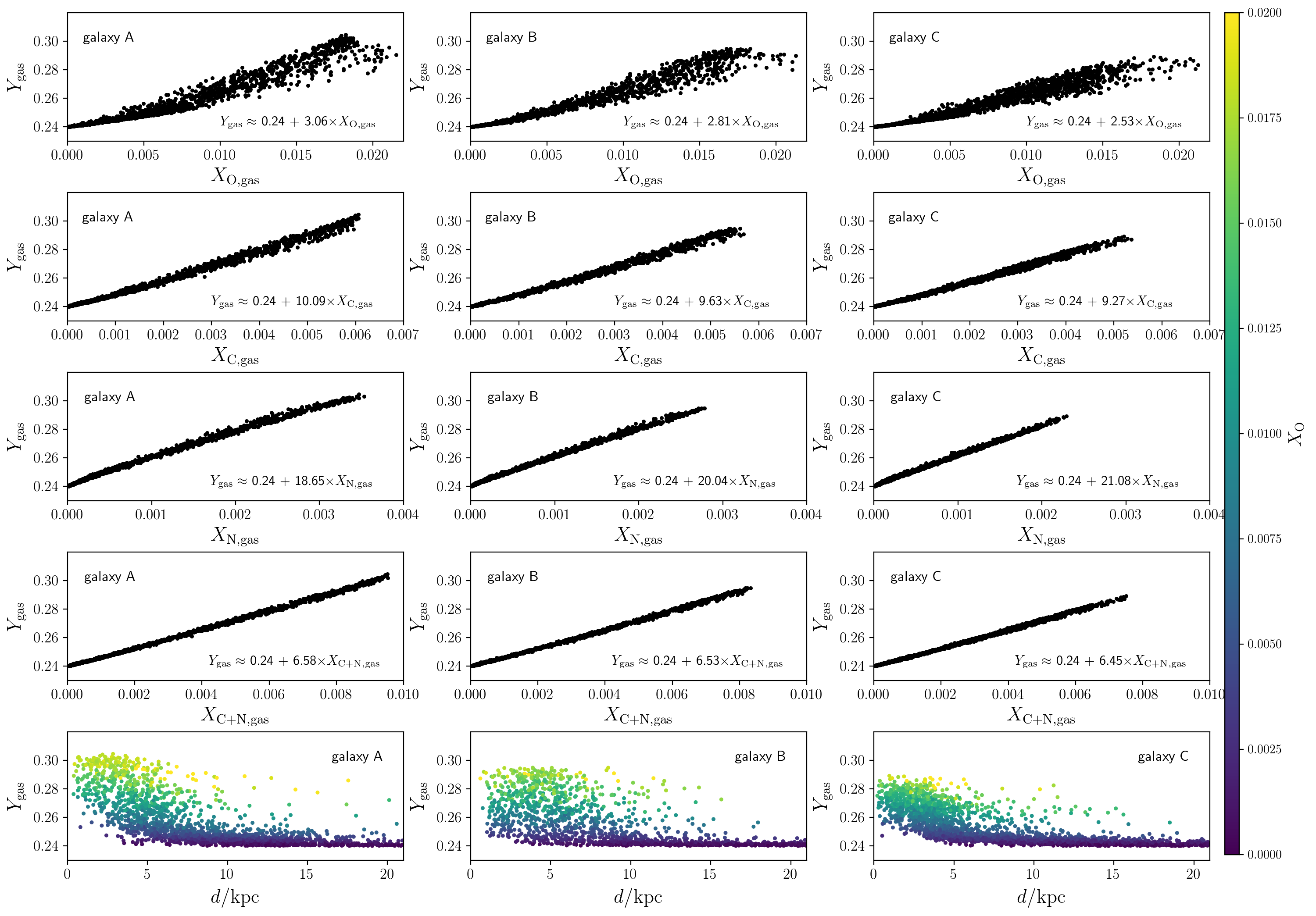} 
\caption{From left to right, the various columns correspond to each of our reference galaxy, from galaxy A to galaxy C, respectively. In the first row, at the top, we show how the gas-phase He mass fraction, $Y_{\text{gas}}$, varies as a function of the gas-phase O mass fraction, $X_{\text{O,gas}}$; in the second row, we show $Y_{\text{gas}}$ versus $X_{\text{C,gas}}$; in the third row, we show $Y_{\text{gas}}$ versus $X_{\text{N,gas}}$; in the fourth row, we show $Y_{\text{gas}}$ versus $X_{\text{C+N,gas}}$; finally, in the fifth row, at the bottom, we show our predictions for the radial profile of $Y_{\text{gas}}$ as a function of the galactocentric distance, $d$, with the colour coding representing the O mass fraction. 
}
\label{fig1}
\end{figure*}

\subsection{The three reference galaxies}
 In this paper, we investigate how He abundances vary within three simulated disc galaxies, which we have selected out of the ten reference galaxies of \citet{vincenzo2018b}. In particular, our galaxy A corresponds to galaxy 3 of \citet{vincenzo2018b}, galaxy B to galaxy 6, and galaxy C to galaxy 9. Nevertheless, our results for He would have been the same, had we selected other galaxies from \citet{vincenzo2018b}\footnote{These galaxies ABC are different from the sample in \citet{vincenzo2018a}.}.  

 In Fig. \ref{fig0} we show the age distribution of the stellar populations in galaxy A-C as predicted at the present time. 
 The different curves in Fig. \ref{fig0} are normalised such that the area below each of them corresponds to the total galaxy stellar mass at redshift $z=0$.  By looking at the figure, it is clear that -- from galaxy A to galaxy C -- the SFH becomes more and more concentrated towards later epochs of the cosmic time.
 
In summary, our three reference galaxies at redshift $z=0$ have the following total stellar and gas masses: \textit{(i)} $M_{\star,\text{A}} = 3.29\times10^{10}\;\text{M}_{\sun}$ and  $M_{\text{gas},\text{A}} = 1.10\times10^{10}\;\text{M}_{\sun}$; \textit{(ii)} $M_{\star,\text{B}} = 1.60\times10^{10}\;\text{M}_{\sun}$ and  $M_{\text{gas},\text{B}} = 1.21\times10^{10}\;\text{M}_{\sun}$; \textit{(iii)} $M_{\star,\text{C}} = 1.55\times10^{10}\;\text{M}_{\sun}$ and  $M_{\text{gas},\text{C}} = 1.79\times10^{10}\;\text{M}_{\sun}$.


\section{He nucleosynthesis in the simulation} \label{sec:helium}

 Apart from the primordial nucleosynthesis, which accounts for most of the He in the Universe, He can be synthesised and released into the ISM of galaxies by {\it all} mass range of stars with $m_{\star}\gtrsim 1\;\text{M}_\odot$. The main source of uncertainty for the He nucleosynthesis in our cosmological simulation is given \textit{(i)} by the assumptions about stellar rotation and mass loss in massive star models (see the review by \citealt{maeder2009}), and \textit{(ii)} by the treatment of overshooting, convective boundary conditions, hot-bottom burning, and efficiency of the third dredge-up (mixing processes) in AGB stellar models (see the detailed discussion in \citealt{ventura2013,renzini2015,karakas2016,karakas2018}).   
 
 In Fig. \ref{fig:heliumcontribution}, we show how the average IMF-weighted stellar yields of He, $\langle \mathcal{P}_{\text{He}} \rangle$, vary as functions of the initial stellar metallicity, $Z_{\star}$, by assuming some different stellar models for massive and AGB stars. In particular, the quantity $\langle \mathcal{P}_{\text{He}} \rangle$ is computed as follows:
\begin{equation} \label{eqa}
\langle \mathcal{P}_{\text{He}}(Z_{\star}) \rangle_{\text{MS}} =
\int_{ 13\,\text{M}_{\sun} }^{ 40\,\text{M}_{\sun} }{ dm \, \mathrm{IMF}(m) \, p_{\text{He}}(m,Z_{\star}) },
\end{equation}
\noindent for massive stars, where $p_{\text{He}}(m,Z_{\star})$ represents the He stellar yields as functions of mass and metallicity, and
\begin{equation} \label{eqb}
\langle \mathcal{P}_{\text{He}}(Z_{\star})\rangle_{\text{AGB}} =
\int_{ 0.8\,\text{M}_{\sun} }^{ 8\,\text{M}_{\sun} }{ dm \, \mathrm{IMF}(m) \, p_{\text{He}}(m,Z_{\star}) },
\end{equation}
\noindent for AGB stars. 

In Fig. \ref{fig:heliumcontribution}(a), the average stellar yields of He from the massive star models of \citet{nomoto2013} -- which do include mass loss and SN nucleosynthesis but not the effect of rotation -- are compared with the average yields of \citet{limongi2018}, which do include also the effect of rotation. In particular, the different curves in Fig. \ref{fig:heliumcontribution}(a) for the \citet{limongi2018} yields correspond to models with different stellar rotational velocities (v0, v150, and v300, which stand for $v_{\text{rot}}=0$, $150$, and $300\;\text{km}\,\text{s}^{-1}$, respectively). At sub-solar metallicities, massive star models with higher and higher rotational velocities give rise -- on average -- to smaller amounts of He; moreover, the stellar yields of He from rotating stellar models show a stronger dependence on metallicity than non-rotating massive star models; as we will see in Section \ref{subsec:3.2}, these resulting differences in the stellar yields for massive stars give rise to a systematic uncertainty in the predictions of chemical evolution models.  

Finally, in Fig. \ref{fig:heliumcontribution}(b), the average stellar yields of He from the AGB stellar models of \citet{ventura2013} are compared with those of \citet{karakas2010}, assumed in our cosmological simulation. We find that the \citet{karakas2010} stellar yields of He decrease as functions of metallicity, and this makes the \citet{ventura2013} stellar yields larger by a factor of $\approx2$ at solar and super-solar metallicities. The most important differences between the two stellar models reside in the physics of super-adiabatic convection and mixing \citep{karakas2002,karakas2007}, which affect the final stellar yields from AGB stars. 

\subsection{Production rate of He from SSPs of different metallicity in our cosmological simulation}

In order to understand how He is produced by the stars in our simulation, in Fig. \ref{fig:heliunrate}, we show how much He is deposited by simple stellar populations (SSPs) of different metallicity, per unit time and per unit mass of the SSPs (namely, in units of $\text{M}_{\sun}\,\text{Gyr}^{-1}\,\text{M}^{-1}_{\text{SSP}}$, where $\text{M}_{\text{SSP}}$ represents the initial mass of the SSP), by assuming the same stellar lifetimes and IMF as in our cosmological simulation. We remark on the fact that each star particle in our simulation is treated as a SSP \citep{kobayashi2004}. 

\begin{figure}[t!]
\centering
\includegraphics[width=8cm]{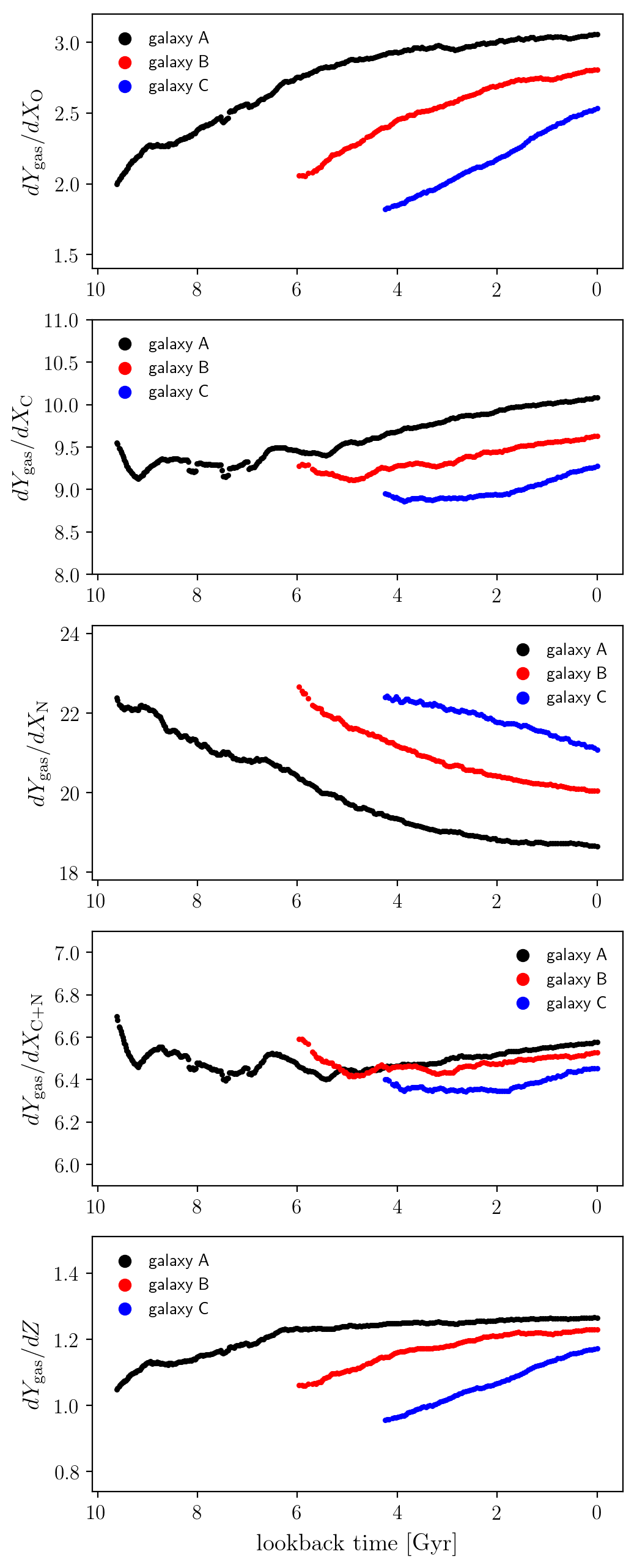} 
\caption{ The predicted evolution of $dY_{\text{gas}}/dX_{\text{O}}$ (top panel), $dY_{\text{gas}}/dX_{\text{C}}$ (second panel), $dY_{\text{gas}}/dX_{\text{N}}$ (third panel), $dY_{\text{gas}}/dX_{\text{C+N}}$ (third panel), and $dY_{\text{gas}}/dZ$ (bottom panel) as a function of the look-back time. The black curve corresponds to galaxy A, the red curve to galaxy B, and the blue curve to galaxy C. The values of $dY_{\text{gas}}/dX_{\text{C,N,O,}Z}$ correspond to our best-fit values for the predicted $Y_{\text{gas}}$-$X_{\text{C,N,O,}Z}$ relations in each galaxy, by assuming a simple linear law at all times. We only show the evolution of the slopes for times when the galaxy stellar mass $M_{\star}>5\times10^{9}\,\text{M}_{\sun}$.}
\label{fig2}
\end{figure}

The contribution from massive stars and AGB stars to the He chemical enrichment is highlighted in Fig. \ref{fig:heliunrate}; in particular, we compare our results with the \citet{nomoto2013} stellar yields for massive stars (coloured curves) with the stellar yields of \citet[grey shaded area]{limongi2018} for rotating massive stars. There is a tendency such that more metal-rich AGB stars release slightly larger amounts of He into the ISM per unit time. Note that the gap in the figure around $30$-$50$ Myr (blue shaded horizontal region) corresponds to stellar masses in the range between $8$ and $10\,\text{M}_{\sun}$ (even though this mass range is quite uncertain), which are not included in most galactic chemical evolution models including ours. Nevertheless, super-AGB stars do not provide a significant contribution to the global chemical enrichment of galaxies (see, for example, \citealt{vangioni2018}, \citealt{prantzos2018}, and Kobayashi et al. in prep.).

We provide a useful formalism to compute the He production rate from a SSP of age $t$ and metallicity $Z$, by firstly defining a \textit{partial yield per stellar generation} as follows:
\begin{equation} \label{eq1}
\langle \mathcal{Y}_{\text{He}}(t,Z) \rangle = \int_{ m_{\text{TO}}(t,Z) }^{m_{\mathrm{max}}}{ dm \, \mathrm{IMF}(m) \, p_{\text{He}}(m,Z) }, 
\end{equation}
where $p_{\text{He}}$ represents the stellar nucleosynthetic yield of He from all the stars with mass $m$ and metallicity $Z$ in the SSP, as weighted with the assumed IMF, $m_{\mathrm{max}}$ corresponds to the maximum stellar mass which is formed in the SSP, and $m_{\text{TO}}(t,Z)$ represents the turn-off mass (as computed from the inverse stellar lifetimes). The extremes of the integral 
in equation \ref{eq1} enclose all the stars that, at the time $t$ from the birth time of the SSP, have already died, enriching the galaxy ISM with He.

The production rate of He by a SSP with metallicity $Z$ and age $t$ can then be computed as follows:
\begin{equation} \label{eq2}
\frac{d \mathcal{P}_{\text{He}}(t,Z)}{dt} = \frac{  \langle \mathcal{Y}_{\text{He}}(t,Z) \rangle -  \langle \mathcal{Y}_{\text{He}}(t - dt,Z) \rangle }{  dt } 
\end{equation}
which is the quantity shown on the $y$-axis of Fig. \ref{fig:heliunrate}.

\begin{figure}[t!]
\centering
\includegraphics[width=8cm]{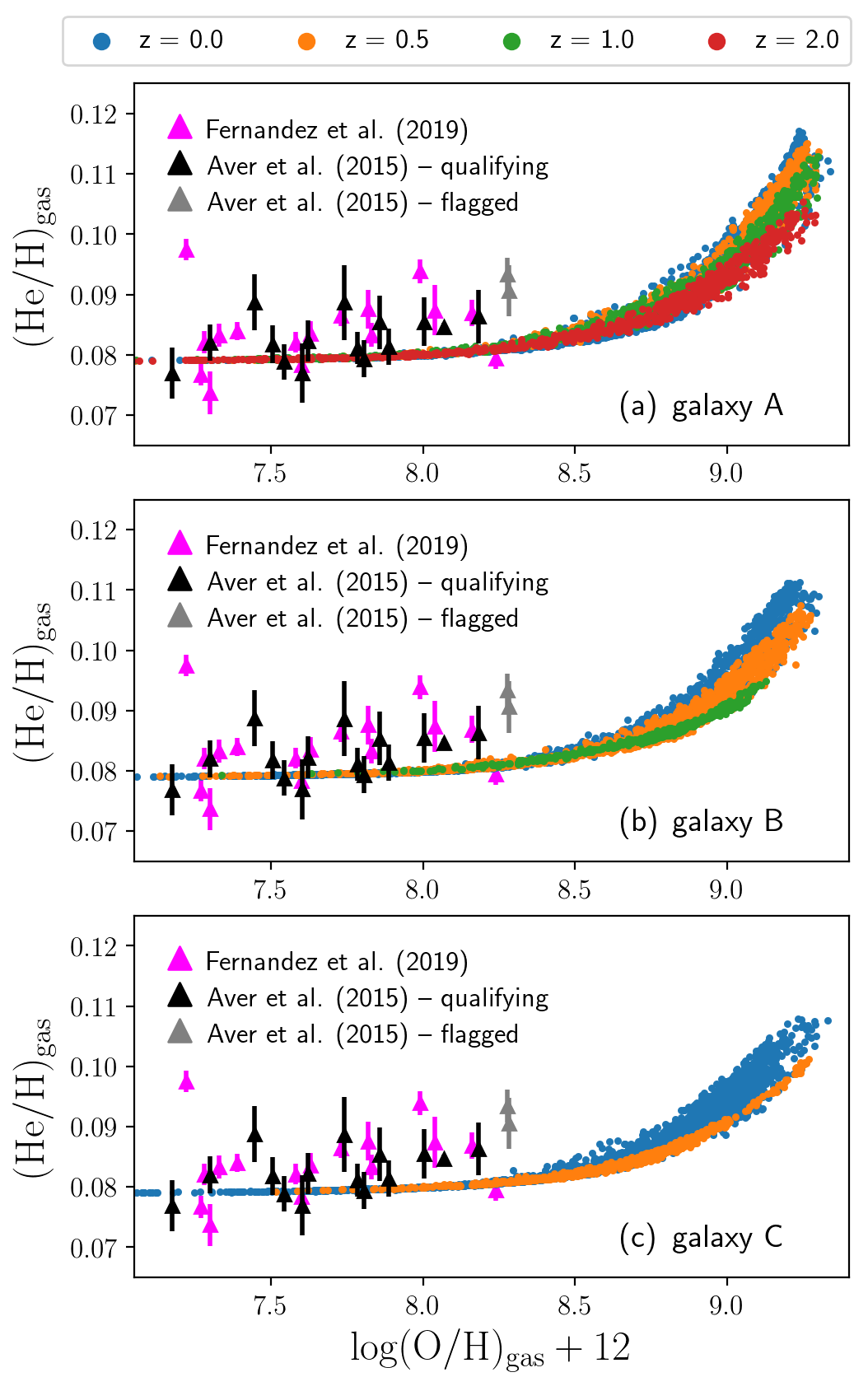} 
\caption{
The redshift evolution of the gas-phase He/H-O/H abundance pattern in our three simulated disc galaxies. The blue points correspond to the predicted gas-phase abundances at redshift $z=0$, orange points to $z=0.5$, green points to $z=1$, and red points to $z=2$. The black and gray triangles with error bars correspond to the observational data of \citet{aver2015}, for a sample of metal-poor HII regions within 16 emission-line galaxies in the Local Universe; the black triangles correspond to their qualifying sample, whereas the gray triangles mark the abundances affected by large systematic uncertainty. Finally, the magenta points with error bars correspond to the observational data of \citet{fernandez2019}. }
\label{fig:heh_redshiftevolution}
\end{figure}

\begin{figure}[t!]
\centering
\includegraphics[width=8cm]{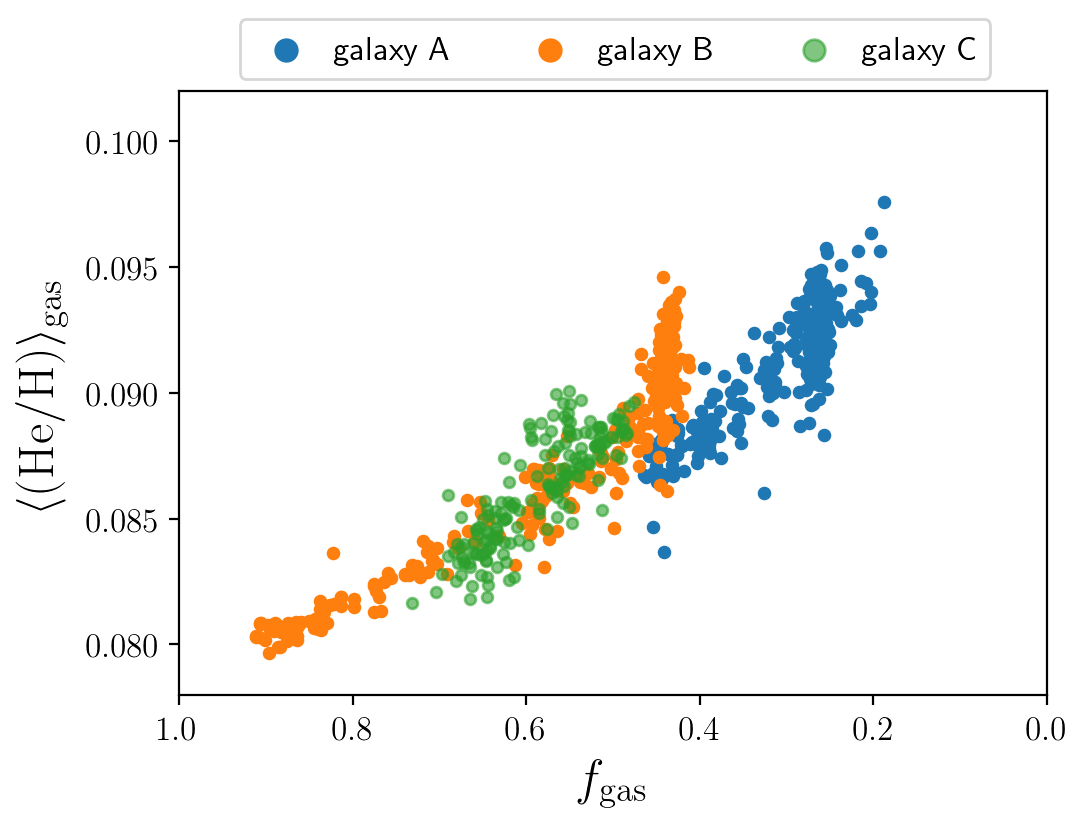} 
\caption{
The evolution of the average SFR-weighted He/H abundances as functions of the gas fraction ($f_{\text{gas}} = M_{\text{gas}}/( M_{\text{gas}} + M_{\star} )$) within our simulated disc galaxies. Blue points correspond to galaxy A, orange points to galaxy B, and green points to galaxy C.
}
\label{fig:heh_fgas}
\end{figure}

\subsection{Quantifying the systematic uncertainty in the predicted Y-Z relation due to different sets of stellar yields} \label{subsec:3.2}

To understand the effect of different stellar yield assumptions for massive stars and AGB stars on the $Y$ versus $Z$ relation in galaxies, in Fig. \ref{fig:tracks} we show the predictions of different one-zone chemical evolution models, assuming a star formation efficiency $\text{SFE}=2.0\,\text{Gyr}^{-1}$, infall time-scale $\tau_{\text{inf}}=4.0\,\text{Gyr}$, and no galactic wind. The models are developed by assuming that the galaxy assembles by accreting a total amount of gas $M_{\text{inf}} \approx 3.16 \times 10^{11}\,\text{M}_{\sun}$, with primordial chemical composition. The accretion of gas into the galaxy potential well follows the following law: $\mathcal{I}(t) \propto e^{-t/\tau_\text{inf}}$. Finally, for the SFR, we assume a linear Schmidt-Kennicutt relation, namely $\text{SFR}(t) = \text{SFE} \times M_{\text{gas}}(t)$.  

In Fig. \ref{fig:tracks}(a,c), we fix the stellar yields of AGB stars and vary the stellar yields of massive stars. In particular, in Fig. \ref{fig:tracks}(a), we fix the stellar yields of \citet{karakas2010} for AGB stars, whereas in Fig. \ref{fig:tracks}(c) we fix for AGB stars the stellar yields of \citet{ventura2013}. The stellar yields of massive stars that we explore in Fig. \ref{fig:tracks}(a,c) are those of \citet{nomoto2013} and those of \citet{limongi2018} for different rotational velocities ($v = 0$, $150$, and $300\,\text{km}\,\text{s}^{-1}$) and iron abundances ($\text{[Fe/H]}=0$, $-1$, $-2$, and $-3\,\text{dex}$) of the progenitor massive star. Finally, in Fig. \ref{fig:tracks}(b,d), we fix the stellar yields of massive stars, and we explore the effect of varying the stellar yields of AGB stars, assuming the yields of \citet{karakas2010} and \citet{ventura2013} for AGB stars. In particular, in Fig. \ref{fig:tracks}(b), we fix the massive star yields of \citet{limongi2018}, whereas in Fig. \ref{fig:tracks}(d) we fix those of \citet{nomoto2013}.

We find that the uncertainty on the predicted $Y$ versus $Z$ relation due to different stellar yield assumptions for massive stars increases as a function of metallicity, being $\Delta Y_{\text{MS}}(0 < Z \la 0.01 ) \sim 0.004$; at super-solar metallicities, where stellar yields are more uncertain, we find $\Delta Y_{\text{MS}}(Z > 0.01) \sim 0.008$. If we fix the stellar yields of massive stars and vary the nucleosynthetic assumptions for AGB stars, we find an uncertainty which, again, increases with metallicity, being $\Delta Y_{\text{AGB}}(0 < Z \la 0.01 ) \sim 0.002$ and $\Delta Y_{\text{AGB}}(Z > 0.01 ) \sim 0.0065$, giving rise to an uncertainty in the predicted $dY/dZ$ due to different stellar yield assumptions for AGB stars, which can be as large as $\approx0.35$.


\section{Results} \label{sec:results}

In this Section we present the results of our analysis for the evolution of the He abundances in our simulated disc galaxies. In Section \ref{subsection1} we focus on the He abundances within the ISM, comparing the predictions of our simulation with the observed He abundances in metal-poor HII regions by \citet{aver2015,fernandez2019}, whereas in Section \ref{subsection2} we study the He content in the stellar populations.  Finally, in Section \ref{subsection-comp} we compare the predictions of our simulation with He abundance measurements in a Galactic open cluster \citep{mckeever2019}, horizontal branch stars in Galactic globular clusters \citep{mucciarelli2014}, RR Lyrae stars in the Galactic bulge \citep{marconi2018}, and in a sample of B-type stars in our Galaxy \citep{morel2006,nieva2012}.

\subsection{He abundances in the ISM} \label{subsection1}

The ISM abundances of He in our simulated galaxies are explored in Fig. \ref{fig1}. From left to right, the three columns of panels in the figure 
represent our three reference galaxies, and -- from top to bottom -- we show our predictions for the following relations in the ISM of the simulated galaxies: \textit{(i)} $Y_{\text{gas}}$ versus $X_{\text{O,gas}}$, \textit{(ii)} $Y_{\text{gas}}$ versus $X_{\text{C,gas}}$, \textit{(iii)} $Y_{\text{gas}}$ versus $X_{\text{N,gas}}$, \textit{(iv)} $Y_{\text{gas}}$ versus $X_{\text{C+N,gas}}$, and \textit{(v)} $Y_{\text{gas}}$ versus galactocentric distance, where $Y_{\mathrm{gas}} = M_{Y,\mathrm{gas}}/M_{\mathrm{gas}}$ represents the He mass fraction within each gas particle in the galaxy, and $X_{\text{O,gas}} = M_{\mathrm{O,gas}}/M_{\mathrm{gas}}$ represents the O mass fraction (similar relations stand also for C, N, and C+N). 


\begin{figure}[t!]
\centering
\includegraphics[width=7.5cm]{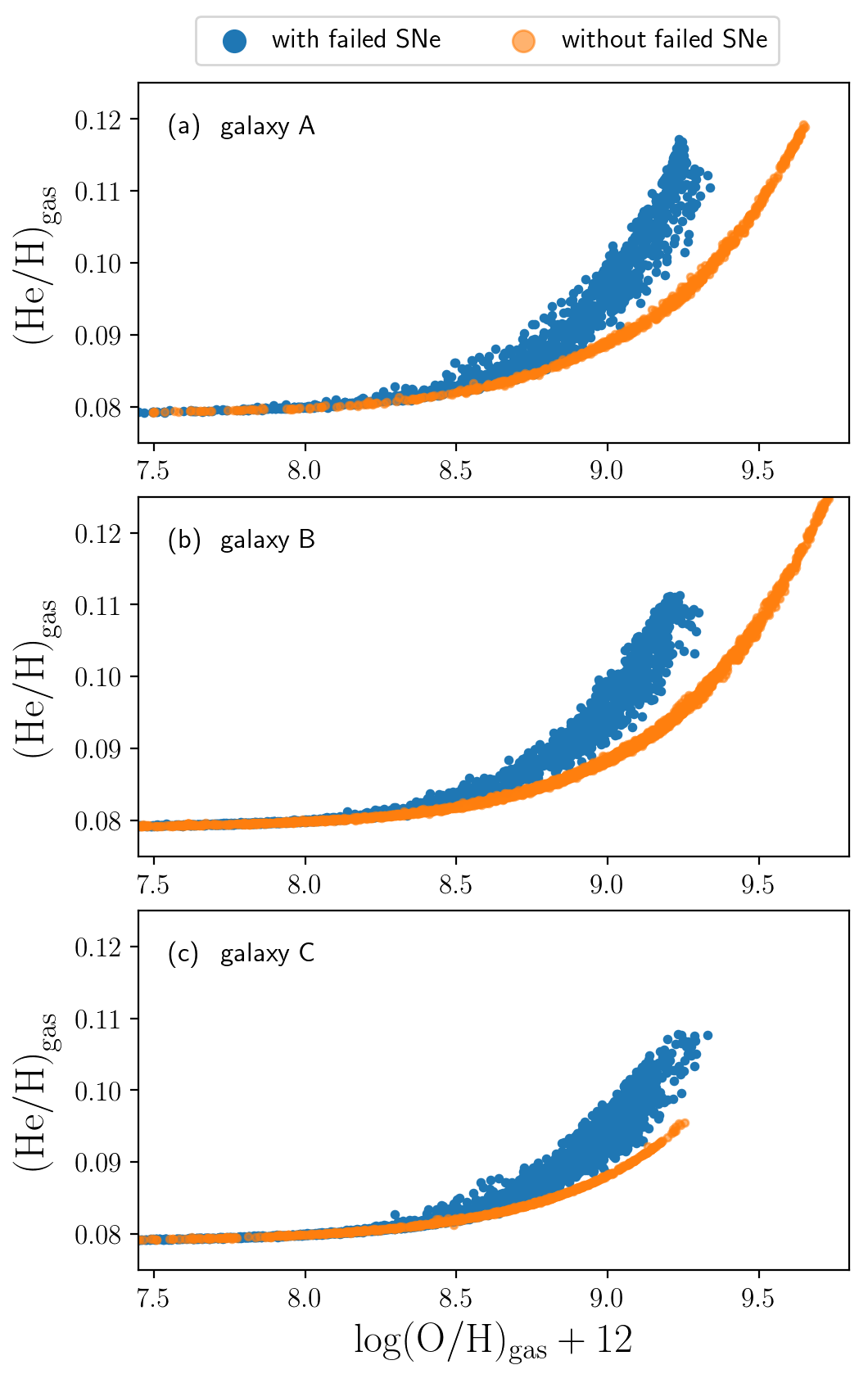} 
\caption{
The predicted gas-phase He/H-O/H abundance pattern at redshift $z=0$ in our three simulated disc galaxies, by assuming failed SNe (blue points), like we do in our reference cosmological hydrodynamical simulation, and by running a similar simulation without failed SNe (orange points) (see also \citealt{vincenzo2018a} for more details). 
}
\label{fig:failed_sne}
\end{figure}

In the lower right of each panel in Fig. \ref{fig1}, we report our best fits to the predicted  $Y$-$X_{\mathrm{O}}$,  $Y$-$X_{\mathrm{C}}$, $Y$-$X_{\mathrm{N}}$, and $Y$-$X_{\text{C+N}}$ relations in our three reference galaxies, by assuming a simple linear law. First of all, as we consider galaxies with SFHs concentrated towards later and later epochs (namely, by moving from galaxy A to galaxy C), we note that the slopes of $Y$-$X_{\mathrm{O}}$ and $Y$-$X_{\mathrm{C}}$ diminish, whereas the slope of $Y$-$X_{\mathrm{N}}$ increases. Secondly, the spread in $Y$-$X_{\mathrm{C,N}}$ is much smaller than that in $Y$-$X_{\text{O}}$. 

The lower scatter in $Y$-$X_{\mathrm{C,N}}$ is a consequence of the fact that the absolute abundances of C and N are much lower than those of O; in particular, we obtain the following approximate value for the relative abundance variations of $i =$ C, N, O, and C+N in our three simulated galaxies, for He abundances in the range $[0.27,0.29]$: 
\begin{equation}
    \frac{ \sigma_{X_{i}} }{ \bar{ X_{i} } } \approx 0.15
\end{equation} 

Our findings suggest that the calibration of the $Y$-$Z$ relation for stellar models should be carried out using nitrogen or carbon abundances; however, since C and N abundances at the stellar surface may be strongly affected by (extra)mixing (e.g., \citealt{iben1983,shetrone2019}), we suggest to use C+N, which is instead a conserved quantity at the stellar surface, as the stars experience dredge-up episodes after they leave the main sequence.

We predict radial gradients of both $Y$ and $X_{O}$ in the ISM of our simulated galaxies (see the bottom panels of Fig. \ref{fig1}); in particular, we find that the most central galaxy regions have higher average metallicites and also higher He abundances than the outermost regions. To reach those high ISM metallicities in the central regions, numerous generations of stars should have succeeded each other polluting the ISM with metals and He, giving rise to very short chemical enrichment time-scales in the galaxy centre. In fact, we find that our simulated disc galaxies grow from the inside out (see also \citealt{vincenzo2018b}), giving rise to chemical enrichment time scales which increase as functions of the galactocentric distance, $d$; this, in turn, determines the predicted trend of $Y$ and $Z$ as functions of $d$.

In Fig. \ref{fig2} we show how our best-fit values for $dY/dX_{\text{O}}$, $dY/dX_{\text{C}}$, $dY/dX_{\text{N}}$, $dY/dX_{\text{C+N}}$, and $dY/dZ$ in the ISM of our three reference galaxies evolve as functions of the look-back time, where $Z$ represents the sum of the abundances of all $31$ chemical elements contributing to metallicity, which are traced in our simulation. We only show our predictions for the epochs when the total galaxy stellar mass is $>5\times10^{9}\,\text{M}_{\sun}$, in order to have an enough number of resolution elements for each galaxy. 
It is clear from the figure that galaxies with different SFHs exhibit also different temporal evolution of $dY/dX_{\text{C,N,O}}$. Interestingly, the temporal evolution of $dY/dX_{\text{N}}$ shows an opposite trend with respect to that of $dY/dX_{\text{O}}$, $dY/dX_{\text{C}}$, and $dY/dZ$.

On the one hand, we predict that $dY/dX_{\text{N}}$ diminishes as a function of time; this is due to the fact that N is mostly produced as a secondary element, and its stellar yields steadily increase with metallicity, whereas the He stellar yields have a weaker dependence on metallicity than those of N. This way, the variation in N between two consecutive time-steps is always larger than that of He, at any epoch of the galaxy evolution. 

On the other hand, $dY/dX_{\text{O}}$ increases with time, because -- in the declining phase of the galaxy SFH -- the variation of He between two consecutive time-steps is larger than that of O; this is due to the large production of He from AGB stars of all masses and metallicities, which pollute the ISM over a large range of delay times from the star formation event. Finally, the evolution of $dY/dX_{\text{C}}$ is weaker than that of $dY/dX_{\text{O}}$ and $dY/dX_{\text{N}}$, because He and C are stricly coupled from the point of view of the stellar nucleosynthesis. Interestingly, the evolution with time of $dY/dX_{\text{C+N}}$, as well as its dependence on the galaxy SFH is relatively weak, varying from $\sim6.4$ to $6.6$.

Our predicted values for $dY/dX_{\text{O}}$ in Fig. \ref{fig2} are consistent with the He abundance measurements in extragalactic HII regions, which use O lines to estimate the ISM metallicity, finding $dY/dX_{O} \sim 2.2$-$2.4$ \citep{izotov2007}. On the other hand, our predicted values for $dY/dZ$ are lower than those determined with indirect He abundance measurements in Galactic open clusters, which report $dY/dZ\sim 1.4$ \citep{brogaard2012}. 

The offset between model and data for $dY/dZ$  can be due to the systematic uncertainty in the He nucleosynthesis from AGB stars. For example, by assuming the \citet{ventura2013} stellar yields, there is a more pronounced He enrichment at high metallicities from AGB stars, giving rise to steeper $Y$-$Z$ relations (i.e. higher $dY/dZ$) than those predicted with the \citet{karakas2010} stellar yields (see Fig. \ref{fig:tracks}). Nevertheless, various studies in the past estimated $dY/dZ \sim 2.1$ for relatively large samples of stars in our Galaxy \citep{jimenez2003,casagrande2007,portinari2010}, corresponding to a value which is much larger than that found in Galactic open clusters by \citet{brogaard2012}. There is, therefore, some uncertainty also on the observational side on the value of $dY/dZ$, mostly because of the indirect methods employed to determine the He content of the stars, which are strongly dependent on the assumptions of stellar evolution models (e.g., \citealt{portinari2010}). 

\subsection{Redshift evolution of the He/H abundances in the ISM}

In Fig. \ref{fig:heh_redshiftevolution}(a-c), we show the predicted redshift evolution of the gas-phase He/H versus O/H abundance pattern in our three simulated disc galaxies. For comparison, we also show the He/H abundance measurements as determined in the HII regions of a sample of 16 metal-poor dwarf irregular galaxies in the local Universe by \citet{aver2015}; the black triangles with error bars correspond to the final qualifying sample of \citet{aver2015}, whereas the grey triangles with error bars correspond to their flagged HII region abundances, which are affected by large systematic uncertainties. Finally, in the same figure, we also show the He/H abundance measurements of \citet[magenta triangles with error bars]{fernandez2019} for a sample of young metal-poor HII regions (see also \citealt{fernandez2018}, for more information about their galaxy sample), which employed a Bayesian approach to fit the spectra in the abundance analysis determination.

\begin{figure}[t!]
\centering
\includegraphics[width=8cm]{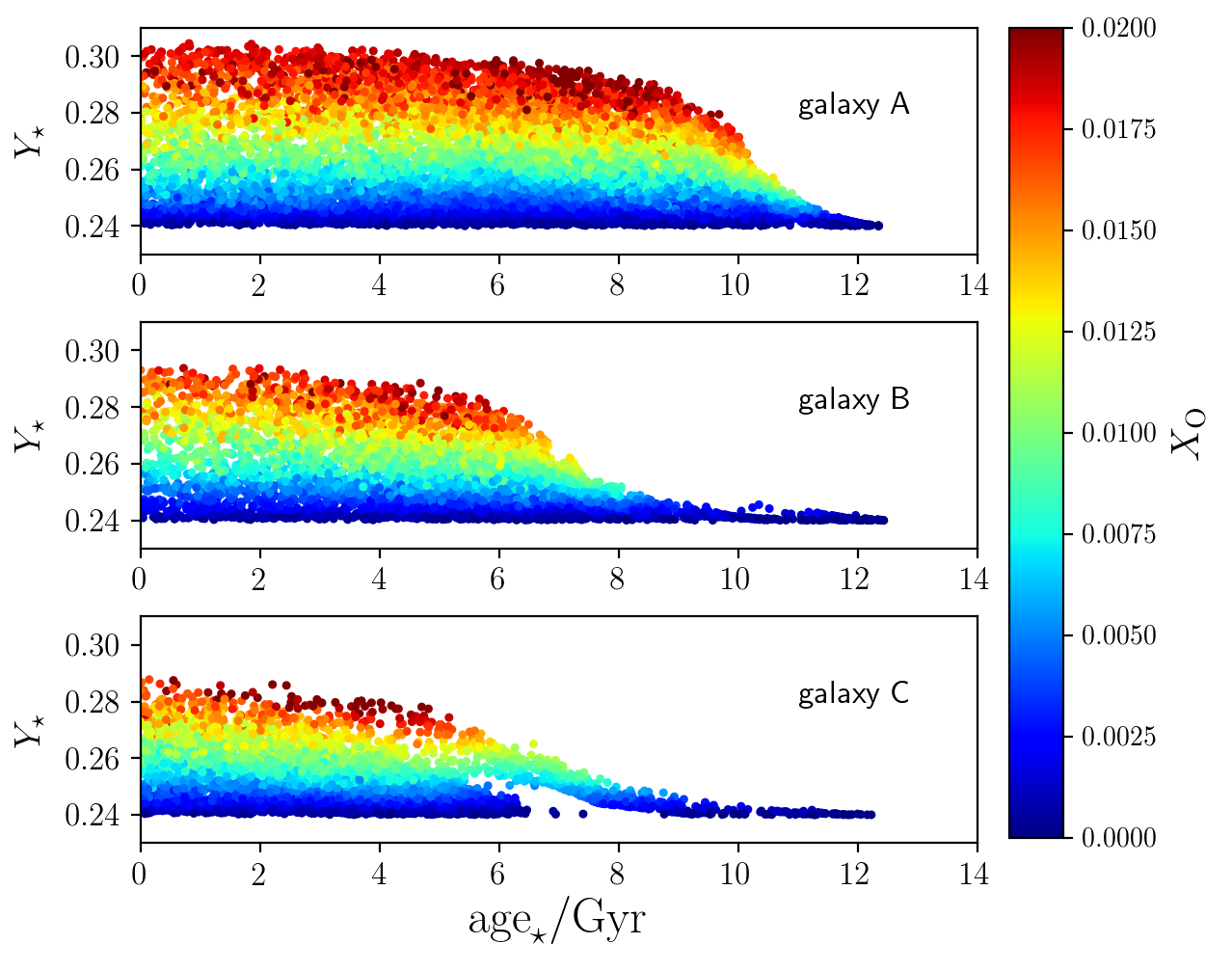} 
\caption{ The predicted He mass fraction in the stellar populations of our three reference galaxies as a function of the stellar ages. The colour coding corresponds to the O mass fraction in the stars. Each panel, from top to bottom, shows our predictions for each reference galaxy, from galaxy A to galaxy C, respectively.  
}
\label{fig3}
\end{figure}

\begin{figure}[t!]
\centering
\includegraphics[width=8cm]{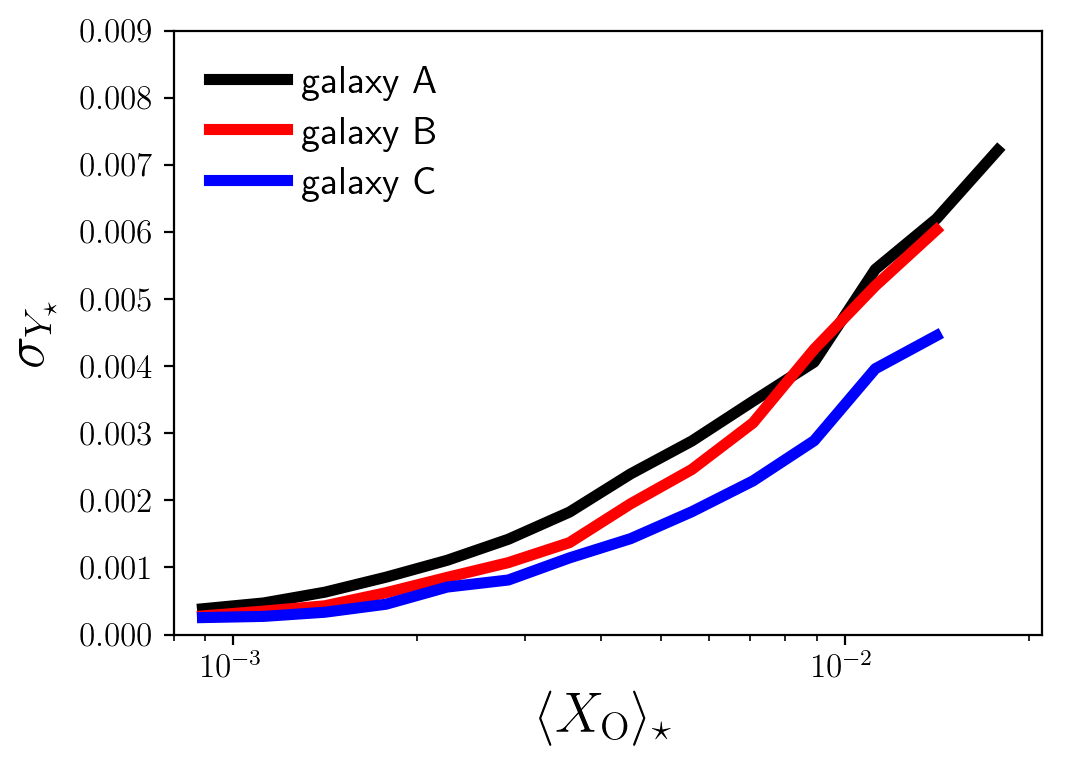} 
\caption{ The dispersion of the He abundances in the stellar populations of our simulated disc galaxies, $\sigma_{Y_{\star}}$, by considering different metallicity bins, $\langle X_{\text{O}} \rangle_{\star}$, with width $0.1\,\text{dex}$ in logarithmic units, as obtained from the analysis of Fig. \ref{fig3}. 
}
\label{fig3b}
\end{figure}

 \begin{figure}[t!]
\centering
\includegraphics[width=8cm]{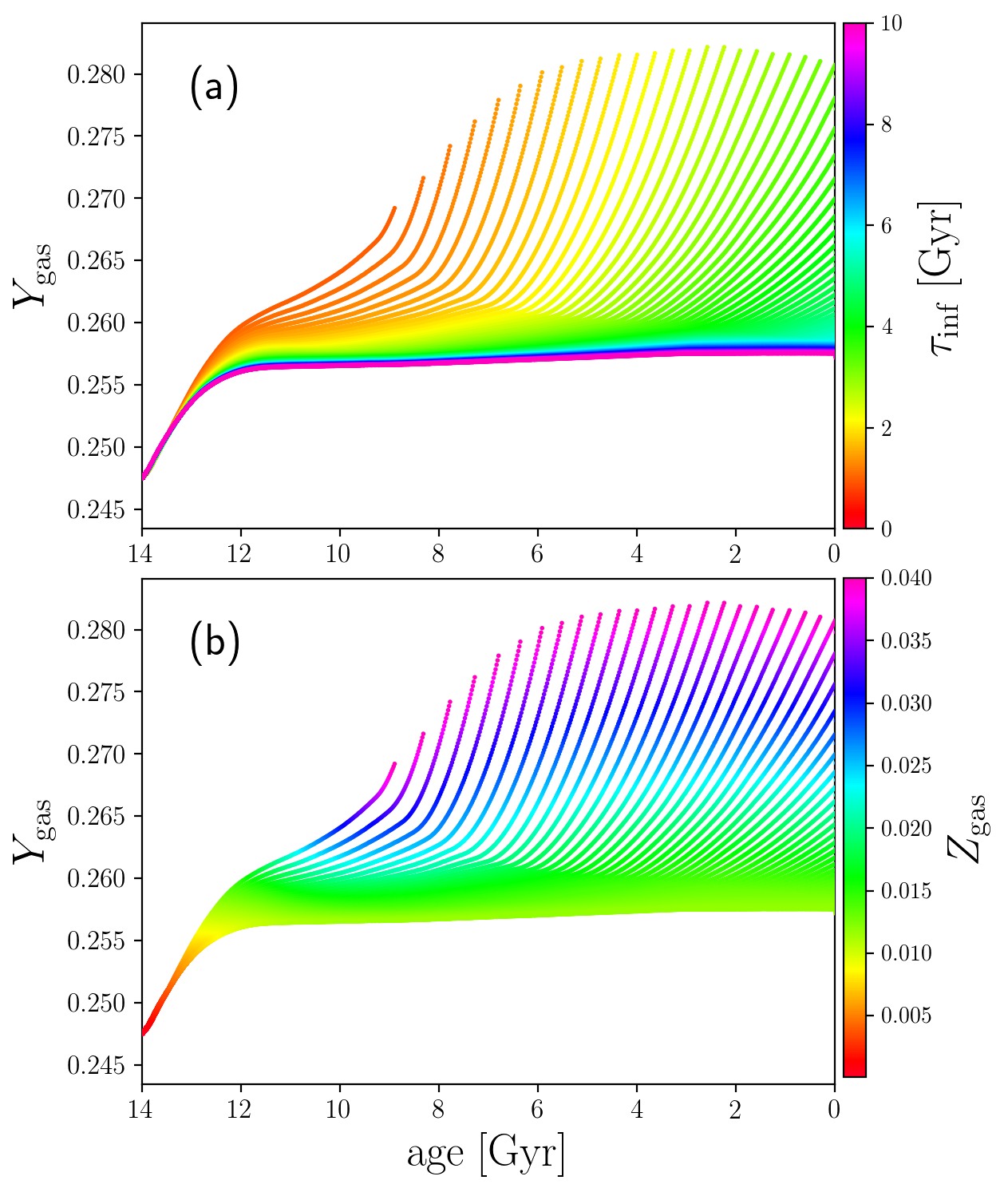} 
\caption{In this figure, we present a set of one-zone chemical evolution models assuming the same IMF, stellar lifetimes and stellar nucleosynthetic yields as in our cosmological hydrodynamical simulation. In the chemical evolution models, we assume a star formation efficiency $\text{SFE}=2\,\text{Gyr}^{-1}$, no galactic winds, and we vary the infall timescale from $\tau=1$ to $10\,\text{Gyr}$, with a step of $0.1\,\text{Gyr}$, to understand whether the inside-out growth of galaxies has an impact on the predicted He abundances. In panel (a), the colour coding represents the infall timescale, whereas in panel (b) the colour coding represents the gas-phase metallicity.  
}
\label{fig:vartau}
\end{figure}

At fixed O/H abundances, we predict that He/H steadily increases as a function of time; such increase of He/H is faster as we move towards higher metallicities, which typically correspond to the more central galaxy regions. This is due to the fact that the most central galaxy regions had the highest star formation activity in the past, giving rise to an enhancement of the He abundances in those regions at the present time. In fact, the SFR in our simulated disc galaxies propagates from the inside out (see also \citealt{vincenzo2018b}), being stronger at the beginning in the galaxy central regions, and reaching on longer typical time scales the outer disc; this explains why fixed He/H abundances are reached at later times, if their corresponding O/H abundances are lower.

\begin{figure*}[t!]
\centering
\includegraphics[width=18cm]{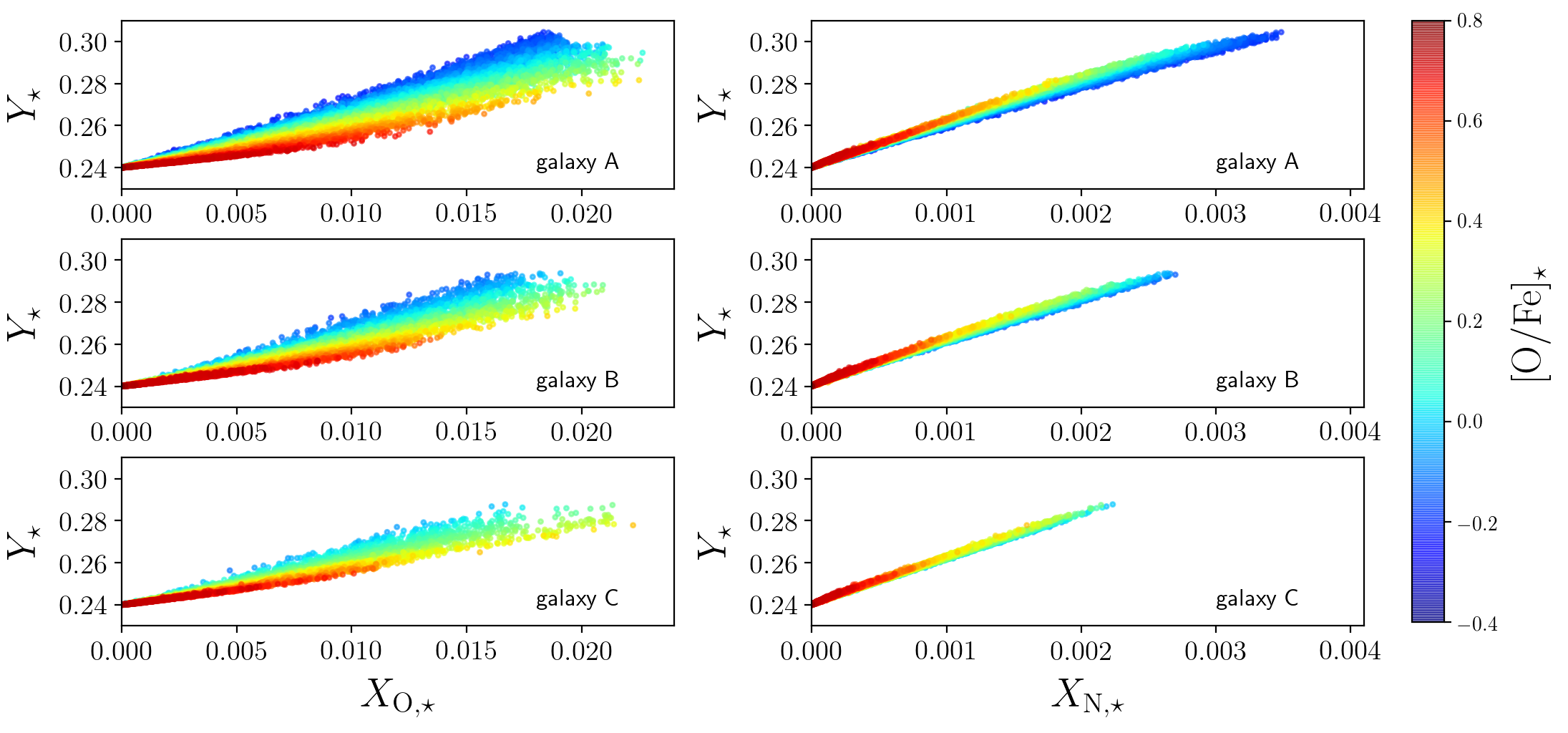} 
\caption{The predicted $Y_{\star}$-$X_{\text{O},\star}$ (left panels) and $Y_{\star}$-$X_{\text{N},\star}$ (right panels) in the stellar populations of our three reference galaxies (from top to bottom). The colour coding represents the $[\alpha/\text{Fe}]_{\star}$ ratios of the stars. }
\label{fig4}
\end{figure*}

\begin{figure}[t!]
\centering
\includegraphics[width=8cm]{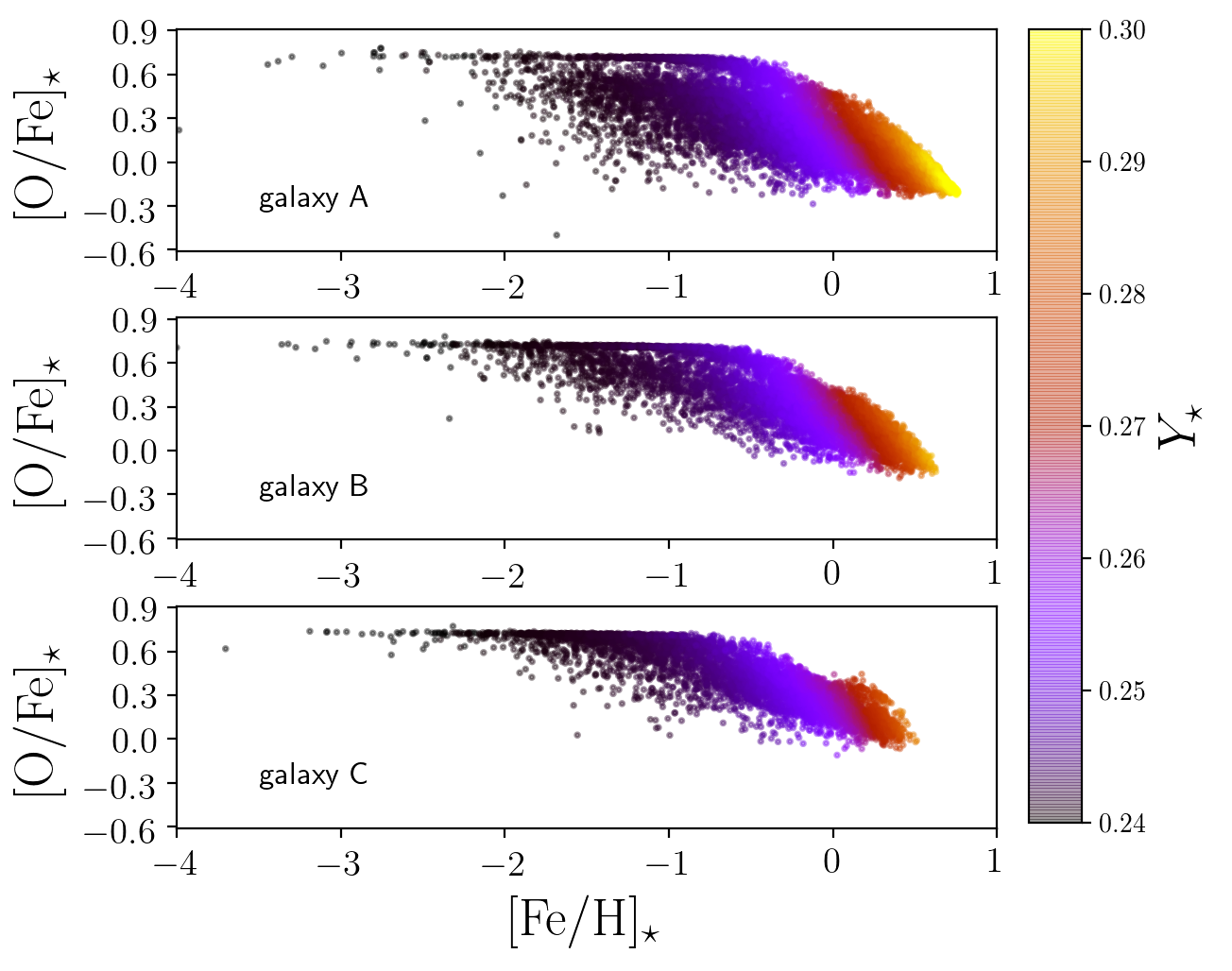} 
\caption{The predicted $[\mathrm{O/Fe}]_{\star}$-$[\mathrm{Fe/H}]_{\star}$ diagrams in the stellar populations of our three reference simulated galaxies, with the colour coding representing the He content, $Y_{\star}$, of the stars. }
\label{fig5}
\end{figure}

We remark on the fact that the observed data in Fig. \ref{fig:heh_redshiftevolution}(a-c) correspond to abundances within \textit{different} metal-poor emission-line galaxies, which are put all together in the figure, whereas our simulation data correspond to abundance variations in the ISM of a \textit{single} disc galaxy. Nevertheless, the observations show much more scatter at low metallicity than our simulated disc galaxies, which are characterised -- in their low-metallicity outskirts -- by very homogeneous He/H abundances, about $\approx 0.08$, approximately corresponding to the assumed primordial He/H ratio in the simulation.

The difference in the scatter at low-metallicity between the observations and our simulation may be due to the fact that, in the outskirts of our simulated galaxies, we do not see strong star formation activity; on the other hand, in the observed galaxy samples of \citet{aver2015} and \citet{fernandez2019}, there are prominent H$\alpha$ lines, which are the signature of ongoing star formation activity. These relatively high SFRs at low metallicities may explain such variation in He/H in the observed data. Nevertheless, we remark on the fact that the mass-resolution of our simulation is about $\approx10^{6}\,\text{M}_{\sun}$ for primordial gas particles; therefore, at low metallicity, the pollution of He from a single star formation episode with low intensity would be distributed to a large number of $\sim 10^{6}\,\text{M}_{\sun}$ surrounding gas particles, all with primordial He/H ratio, giving rise to more homogeneous abundances at low-metallicity. Finally, the observed scatter at low metallicity may be the signature of He enrichment from rotating massive Wolf-Rayet stars \citep{kumari2018}, which are not included in our cosmological simulation. 

To demonstrate that the gas fraction is one of the main parameters driving the evolution of the average He abundances in our simulated disc galaxies, in Fig. \ref{fig:heh_fgas}, we show how the average SFR-weighted He/H abundances in the ISM of our three simulated disc galaxies evolve as functions of the galaxy gas fraction, which is defined as $f_{\text{gas}} = M_{\text{gas}}/(M_\text{gas} + M_{\star})$. As the galaxy evolves, because of the continuous star-formation activity, $f_{\text{gas}}$ decreases as a function of time. At the same time, we predict that the average He/H abundances in the gas-phase increase, being He produced by a larger number of ageing stellar populations in the galaxy. 

Finally, in Fig. \ref{fig:failed_sne}, we show how the inclusion of failed SNe affects the predicted He/H versus O/H abundance patterns at redshift $z=0$ in our reference simulated galaxies. Our findings are very similar to those presented in \citet{vincenzo2018a} for the N/O versus O/H abundance pattern; in particular, we find that the inclusion of failed SNe increases the inhomogeneity of the chemical abundance patterns, making also the final galaxies less metal-rich at the present time. The assumption of failed SNe changes the evolution of the metallicity as a function of time, by shifting the stellar metallicity distribution function towards lower metallicities; this way, there is -- on average -- a larger production of He as a function of time (see also Fig. \ref{fig:heliumcontribution}, which shows that AGB stars with higher metallicities eject -- on average -- lower amounts of He per unit time). Therefore, at fixed O/H abundances, the simulation with failed SNe predicts higher He/H abundances in the galaxy ISM.

\subsection{He abundances in the stellar populations} \label{subsection2}

In Fig. \ref{fig3} we investigate how the He abundances in the stars of our simulated galaxies vary as functions of the stellar ages. The colour coding in the figure represents the metallicity of the star particles, as traced by the O abundance. 

Our findings in Fig. \ref{fig3} can be easily interpreted in light of our previous results in Section \ref{subsection1}.
In particular, at any given galacto-centric radius on the galaxy disc, stellar populations of different ages and metallicities cohabit, depending on how the past star formation activity was distributed in space as a function of time. Therefore, for a given age of the stars, there is a distribution of metallicities, which automatically translate into $Y$ variations. At any given bin of stellar ages, the star particles with the lowest metallicities were typically born in the galaxy outermost regions, where -- because of the inside-out growth of the galaxy disc -- the chemical enrichment time scales are typically long. This gives rise to low He abundances in the galaxy outer regions, close to $Y\sim0.24$, which is approximately the primordial value.

To understand the impact of different stellar ages on the predicted He abundances, we divide the stellar populations in Fig. \ref{fig3} in different bins of metallicity, with width $0.1\text{dex}$ in logarithmic units, and -- for each bin -- we compute the dispersion of the He abundances, $\sigma_{Y_{\star}}$, as due to the variation of the stellar ages in the bin. 
Our results are shown in Fig. \ref{fig3b}. We find that the dispersion of the He content due to different stellar ages increases as a function of metallicity, reaching values as high as $\approx 0.007$ for galaxy A, which corresponds to $\sim 12$ per cent of the predicted global variation of $Y$ within the galaxy.

\begin{figure}[t!]
\centering
\includegraphics[width=8cm]{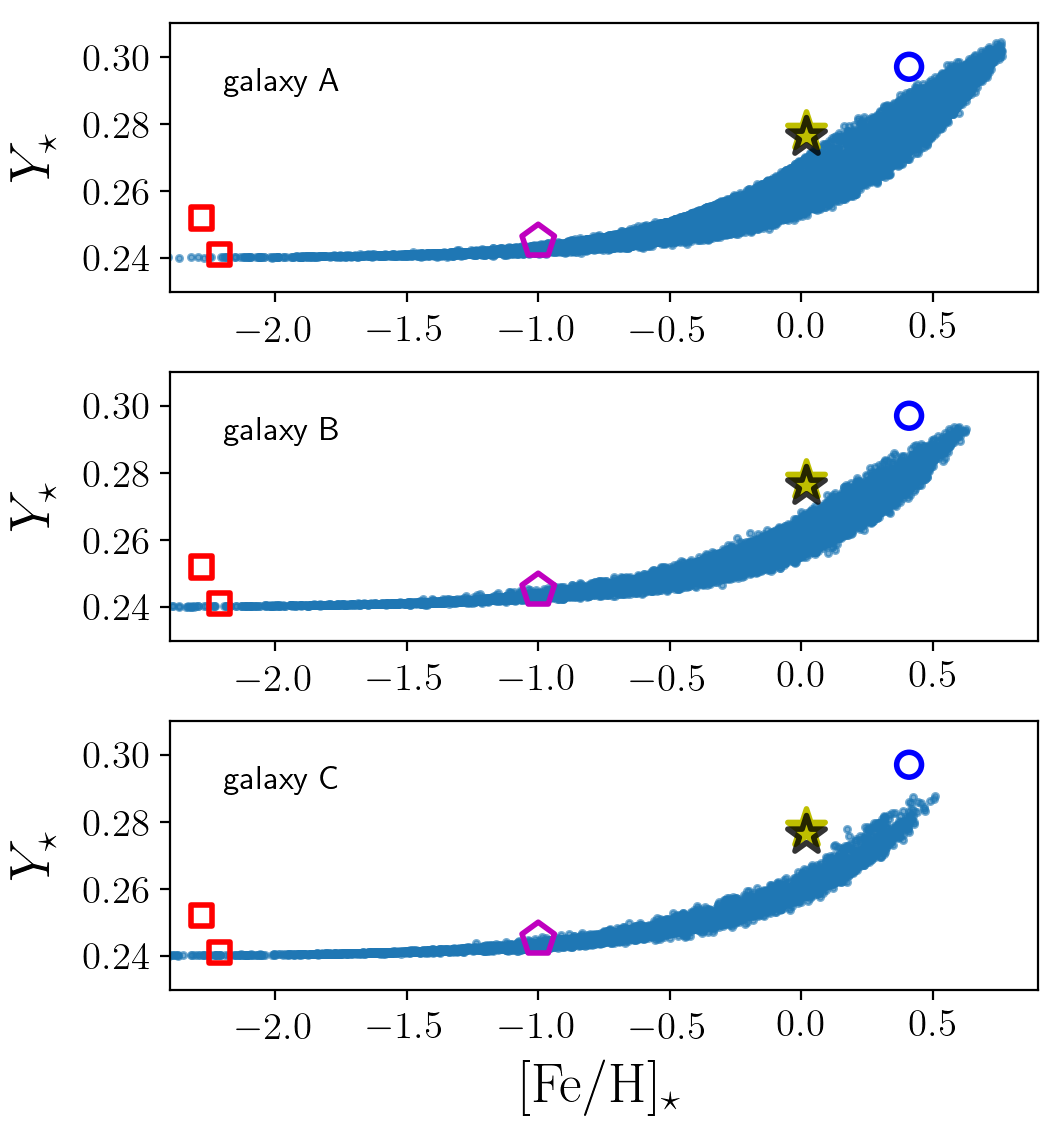} 
\caption{In this figure, we compare our predicted $Y_{\star}$ versus $[\text{Fe/H}]_{\star}$ relations in galaxy A-C (from top to bottom; light blue points) with the observed data of \citet[red squared]{mucciarelli2014} for blue horizontal-branch stars in the Galactic globular clusters M30 and NGC 6397, \citet[magenta pentagon]{marconi2018} for a sample of RR Lyrae in the Galactic bulge, the He abundances in the Galactic open cluster NGC6791 from \citet[blue circle]{mckeever2019}, and the He abundances in a sample of B-type stars within the Solar neighbourhood from \citet[black star]{nieva2012}. The yellow star corresponds to the initial He abundance of the Sun \citep{serenelli2010},
for which $\text{[Fe/H]}$ is computed by assuming that the iron mass fraction scale with respect to the total metallicity like in the solar photospheric chemical composition as derived by \citet{grevesse1998}. }
\label{fig6}
\end{figure}

\begin{figure}[t!]
\centering
\includegraphics[width=8cm]{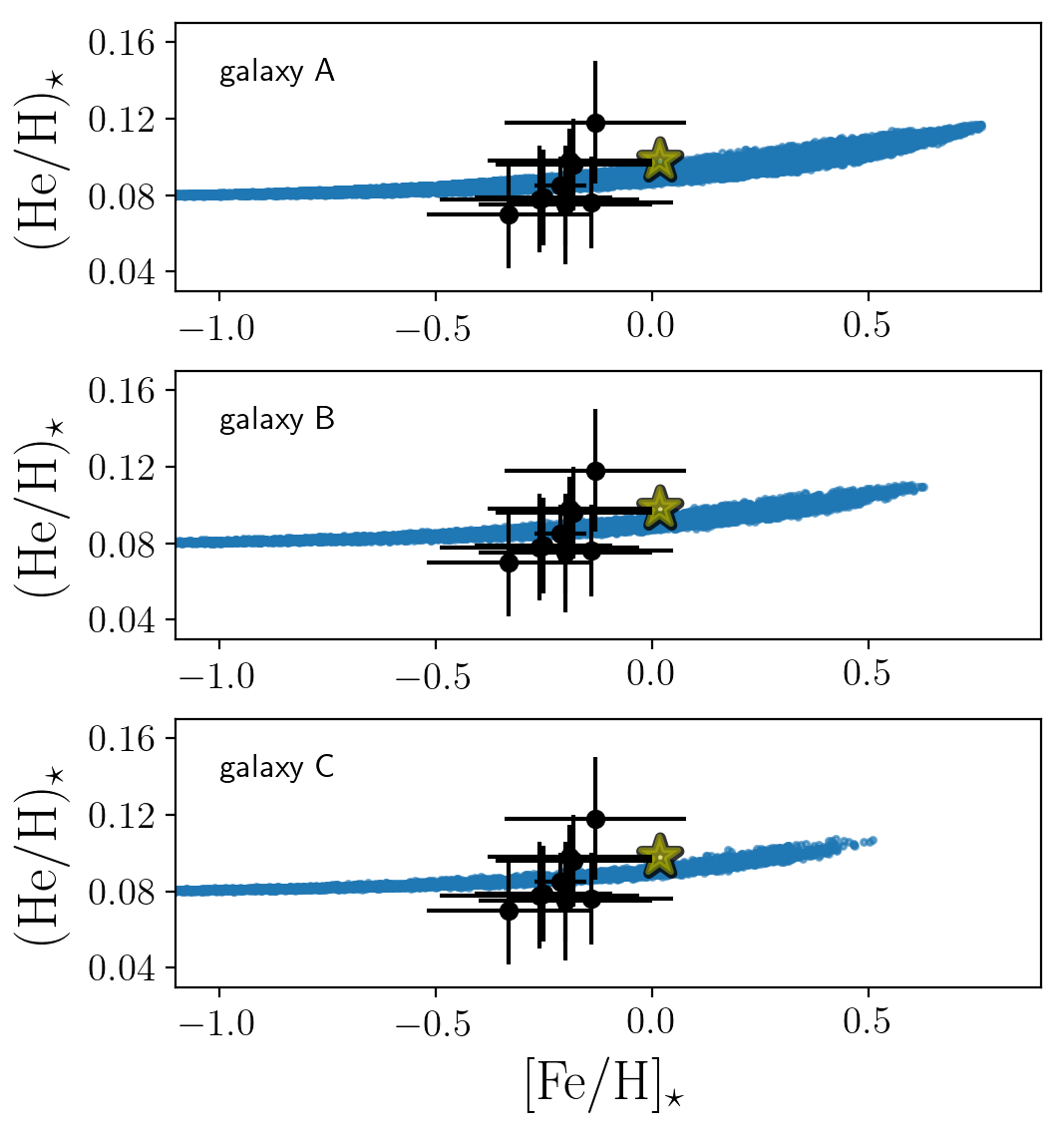} 
\caption{In this figure, we compare our predicted $\text{(He/H)}_{\star}$ versus $[\text{Fe/H}]_{\star}$ relations in galaxy A-C (from top to bottom; light blue points) with the observed data of \citet[black points with error bars]{morel2006} for B-type stars. We also show the value of the typical He abundance in a sample of B-type stars within the Solar neighbourhood from \citet[black star]{nieva2012}. The yellow star corresponds to the initial He abundance in the Sun \citep{serenelli2010}, where we scale the Fe abundances with respect to the solar photospheric abundance of Fe from \citet{grevesse1998}. }
\label{fig7}
\end{figure}

In Fig. \ref{fig:vartau}, we present a set of one-zone chemical evolution models (similar as those described in Section \ref{sec:helium}) assuming the same IMF, stellar lifetimes and stellar nucleosynthetic yields as in our cosmological hydrodynamical simulation. In this set of models, we assume a star formation efficiency $\text{SFE}=2\,\text{Gyr}^{-1}$, no galactic winds, and we vary the infall time scale from $\tau=1$ to $10\,\text{Gyr}$, with a step of $0.1\,\text{Gyr}$, to determine whether the inside-out growth of the galaxy disc has an impact on the predicted $Y$ versus \textit{age} relation. In particular, in the context of the inside-out growth of galaxies, shorter infall time scales correspond to more central galaxy regions. In Fig. \ref{fig:vartau}(a), the colour coding represents the infall timescale, whereas in Fig. \ref{fig:vartau}(b) the colour coding represents the gas-phase metallicity. 

Since the assumed SFE in our chemical evolution models would only systematically shift the $Y$-$Z$ and $Y$-\textit{age} relations along the $y$-axis, making faster or slower the chemical enrichment as a function of time (if one increases or decreases the SFE, respectively), Fig. \ref{fig:vartau} demonstrates that the inside-out growth of the galaxy disc is the main effect regulating the evolution of the He abundances in galaxies. In particular, diminishing the infall time scale (namely, moving towards inner galaxy regions) determines a faster chemical enrichment in the galaxy, giving rise to higher metallicities and He abundances, for a fixed age. Our chemical evolution models, therefore, predict that the innermost galaxy regions (e.g., bulge and inner disc) have steeper $Y$ versus \textit{age} relations at high metallicities than the outermost galaxy regions. 

In Fig. \ref{fig4} we investigate how $Y_{\star}$-$X_{\text{O},\star}$ (left panels) and $Y_{\star}$-$X_{\text{N},\star}$ (right panels) in the stars of our simulated galaxies depend on $[\alpha/\mathrm{Fe}]_{\star}$ (colour coding, where we assume O as a proxy for the $\alpha$-elements). There is a trend, according to which the high-$[\alpha/\text{Fe}]$ stars have flatter $Y_{\star}$-$X_{\text{O},\star}$ and steeper $Y_{\star}$-$X_{\text{N},\star}$ relations than the low-$[\alpha/\text{Fe}]$ stars, even though the predicted spread in $Y_{\star}$-$X_{\text{O},\star}$ is larger than that in $Y_{\star}$-$X_{\text{N},\star}$. These predictions are consistent with our results on the temporal evolution of $dY/dX_{\text{O}}$ and $dY/dX_{\text{N}}$ in the ISM (see Fig. \ref{fig2}, top and bottom panels). 

Finally, since in Fig. \ref{fig4} we show our predictions for the $[\mathrm{O/Fe}]_{\star}$ ratios, we show that our simulation can produce reasonable results also for the classical $[\mathrm{O/Fe}]_{\star}$-$[\mathrm{Fe/H}]_{\star}$ diagram in Fig. \ref{fig5}, where the colour-coding corresponds to the He content of the stars, $Y_{\star}$. Nevertheless, we caution the readers that Fe and O are produced by Type Ia and core-collapse SNe, respectively, on very different typical time scales from the star formation event.

\subsection{Comparison with the observed He abundances in the stars} \label{subsection-comp}

In Fig. \ref{fig6}, we compare the predictions of our simulation for $Y_{\star}$ versus $[\text{Fe/H}]_{\star}$ in galaxy A-C with the observations. In particular, we show the observed He abundances in the Galactic open cluster NGC 6791 \citep{mckeever2019}, in a sample of horizontal branch stars in M30 and NGC 6397 \citep{mucciarelli2014}, as well as in a sample of RR Lyrae in the Galactic bulge \citep{marconi2018}. We also show the He abundances in in a sample of B-type stars within the Solar neighbourhood from \citet{nieva2012}, and the initial He abundance of the Sun from \citet{serenelli2010}. 
Even though some of the observed data in Fig. \ref{fig6} represent indirect measurements of He in stars (apart from the stellar data of \citealt{nieva2012,mucciarelli2014}), and we did not choose our simulated galaxies to reproduce the observed He abundances in the MW, the predicted trend of $Y_{\star}$ versus $[\text{Fe/H}]_{\star}$ qualitatively agrees with the observed trend, even though our model predictions always lie below the observed MW data at high metallicities.

Finally, in Fig. \ref{fig7}, we compare the predictions of our simulation for $\text{(He/H)}_{\star}$ versus $[\text{Fe/H}]_{\star}$ in galaxy A-C   with the observed He abundances in Galactic B-type stars, as measured by \citet{morel2006}. For reference we also show the value of the typical He abundance of B-type stars \citep{nieva2012} and the initial He abundance of the Sun from \citet{serenelli2010}. The observed data in Fig. \ref{fig7} represent direct He abundance measurements in the stars, and we predict a much less scattered relation than in the observed data, even though our predicted values for $\text{(He/H)}_{\star}$ are qualitatively consistent with observations. 

Galactic globular clusters are nowadays known to host multiple stellar populations, which clearly show up both in their observed CMD (particularly when combining passbands with different response to the molecule bands of OH, CN, CH, and NH; see, for example, \citealt{milone2012}) and in their light-element chemical abundance patterns (see, for example, \citealt{gratton2004,prantzos2006,gratton2012,milone2013}). The He content in the different stellar populations of globular clusters can be inferred by means of precise isochrone-fitting analysis; in particular, stars with higher $Y$ are more luminous and tend to have bluer colours, especially at high metallicities \citep{milone2018}. One of the most puzzling results in globular cluster studies is that their latest generations of stars have enhanced He abundances, which can be as high as $Y\approx 0.4$ (e.g., \citealt{dantona2002,dantona2005,piotto2005,piotto2007}); this has also been confirmed with direct spectroscopic He abundance measurements in globular cluster horizontal branch stars \citep{marino2014}. Finally, such He-enhancement has been found to strongly correlate with the cluster mass, being larger in more massive clusters (e.g., \citealt{milone2015}). 

In our cosmological hydrodynamical simulation, the typical mass of the star particles at redshift $z=0$ is in the range $\approx 10^{5}$ - $10^{6}\,\text{M}_{\sun}$, hence of the order of globular cluster mass. Our simulation cannot naturally predict any He spread within the star particles, because -- by construction -- the latter are simple stellar populations, with fixed age and metallicity. Moreover, the appearance of multiple stellar populations in globular clusters would be a sub-resolution process in our simulation, that could be included only through some parametrisation, without emerging from the simulation itself, like -- for example -- our predicted He enhancement in the central, more dense galaxy regions, which is a natural outcome of the cosmological inside-out growth of the galaxy disc as a function of time. Therefore, in conclusion, if we suppose that our star particles represent globular clusters, then our simulation can provide only an average He content between all the underlying stellar populations. 


\section{Conclusions} \label{sec:conclusions}

In this paper, we have shown, for the first time, how He abundances in star-forming disc galaxies evolve as functions of time, chemical composition, and SFH in the context of cosmological chemodynamical simulations \citep{kobayashi2007,vincenzo2018a,vincenzo2018b}. We believe that the results of our study will be of high interest for a wide range of sub-disciplines in stellar physics, in which the assumed calibration between $Y$ and $Z$ represents one of the major sources of systematic uncertainty, being largely unknown. 

Our main conclusions can be summarised as follows. 
\begin{enumerate}

    \item The predicted $Y$-$X_{\text{C,N,O}}$ relations in galaxies depend on their past SFH (see Figures \ref{fig1}-\ref{fig2}). In particular, young galaxies have -- on average -- flatter $Y$-$X_{\text{C,O}}$ and steeper $Y$-$X_{\text{N}}$ relations than the old ones.
    
    \item We find that $dY/dZ$ depends on the galaxy SFH and is not constant as a function of time. Moreover, the temporal evolution of $dY/dZ$ depends on the particular chemical element which is used to trace $Z$
    
    \item The predicted temporal evolution of $dY/dX_{\text{O}}$ in the ISM is opposite with respect to that of $dY/dX_{\text{N}}$ (see Fig. \ref{fig2}). In particular $dY/dX_{\text{O}}$ increases -- on average -- as a function of time, whereas $dY/dX_{\text{N}}$ decreases, because N is mostly synthesised as a secondary element, with its stellar yields strongly increasing as functions of metallicity. Finally, we find that $dY/dX_{\text{C}}$ weakly increases as a function of time, because He and C are strictly coupled from a nucleosynthesis point of view.  Interestingly, $Y$-$X_{\text{C+N}}$ depends very weakly on the galaxy SFH, having values in the range $\approx[6.4, 6.6]$. 
    
    \item The predicted $Y$-$X_{\text{C}}$ and $Y$-$X_{\text{N}}$ relations are steeper and less scattered than $Y$-$X_{\text{O}}$ (see Figures \ref{fig1}-\ref{fig2}). This suggests to use C, N, or C+N as metallicity calibrators for stellar models, instead of the O abundances. 
    
    \item Our predicted values for $dY/dX_{\text{O}}$ are fairly in agreement with the observed relations in extragalactic HII regions \citep{izotov2007}; however, our predicted values for $dY/dZ$ are lower than those found with indirect He abundance measurements in large samples of stars in our Galaxy \citep{jimenez2003,casagrande2007,portinari2010}, as well in Galactic open clusters \citep{brogaard2012}. This is likely due to the large uncertainty in the He nucleosynthesis from AGB stars in our cosmological simulation; for example, by assuming the \citet{ventura2013} stellar yields for AGB stars, we obtain a steeper $Y$-$Z$ relation than that found with the \citet{karakas2010} stellar yields, with a difference in $dY/dZ$ which can be as large as $\approx 0.35$. Nevertheless, the observed $dY/dZ$ values in the stars still suffers from some  uncertainty, because of the typical indirect methods employed to measure He abundances, which depend on the assumptions of stellar models (e.g., \citealt{casagrande2007,portinari2010}).     
    
    \item We predict radial gradients of $Y$ in the ISM of our simulated disc galaxies, according to which the central regions have -- on average -- higher $Y$ and metallicities than the outermost regions (see Fig. \ref{fig1}). This is due to an inside-out growth of the stellar mass (and to an inside-out propagation of the star formation activity) in our simulated disc galaxies as a function of time (see also \citealt{vincenzo2018b}), the main effect of which -- from the point of view of chemical evolution -- is an increase of the typical chemical enrichment time scale as a function of the galactocentric distance, giving rise to higher average $Y$ and $Z$ values in the centre. 
    
    \item We find that, at fixed O/H, the predicted gas-phase He/H abundances increase as a function of time, with such increase being faster in the inner galaxy regions (see Fig. \ref{fig:heh_redshiftevolution}). We conclude that this is an effect of the inside-out growth of our simulated disc galaxies as a function of time.
    
    \item By comparing our simulations with the observed He/H abundances in a sample of low-metallicity star-forming galaxies from \citet{aver2015,fernandez2019}, we predict much more homogeneous gas-phase He/H abundances at low metallicities than in the observed data set (see Fig. \ref{fig:heh_redshiftevolution}). This may be due to the fact that the observed galaxy sample are relatively metal-poor with ongoing star formation activity at the present time, whereas the low-metallicity environments in our simulated disc galaxies at the present time correspond to the galaxy outer regions, where there is no sign of recent strong star formation activity. Nevertheless, we also discuss that this disagreement in the scatter may be due to the limited resolution of our simulation, as well as to chemical enrichment from rotating massive Wolf-Rayet stars \citep{kumari2018}, which are not included in our cosmological simulation.  
    
    \item For a fixed stellar metallicity bin, the variation of $Y_{\star}$ due to different stellar ages becomes more and more important at higher metallicities (see Fig. \ref{fig3b}). Since the stars in the bulge and in the inner disc of galaxies typically have the highest metallicities, we expect that the impact of the stellar ages on the variation of the He abundances becomes more important in the inner galaxy regions; in particular, we find that, for the typical high metallicities of Galactic bulge stars, the variation of $Y_{\star}$ contributed by different stellar ages can be as high as $\sim 12$ per cent with respect to the global variation of $Y_{\star}$. On the other hand, in the galaxy disc, the metallicity is the most important quantity determining the variation of $Y_{\star}$ in the galaxy. Nevertheless, the systematic uncertainty introduced by different stellar yield assumptions for AGB stars is comparable to the spread that we find for $Y_{\star}$ at high metallicity, which is of the order of $\approx 0.007$. 
    
    \item By making use of chemical evolution models with different infall time scales, we find that the innermost galactic regions (which are characterised by shorter infall time scales, according to the inside-out scenario) have steeper $Y$ versus \textit{age} relations at high metallicities than the outermost disc regions (see Fig. \ref{fig:vartau}). 
    
    \item The predicted $Y_{\star}$-$X_{\text{O},\star}$ relation in the stars is very sensitive to $[\alpha/\text{Fe}]$. In particular, the high-$[\alpha/\text{Fe}]$ stars exhibit a flatter average relation than those with low $[\alpha/\text{Fe}]$. An opposite but weaker trend is found for $Y_{\star}$-$X_{\text{N},\star}$, when considering stars with different $[\alpha/\text{Fe}]$ (see Fig. \ref{fig4}). 
    
    \item Even though we did not choose our simulated disc galaxies to reproduce the observed chemical abundances of He in our Galaxy, the predicted trend of our simulations for $Y_{\star}$ versus $[\text{Fe/H}]_{\star}$ qualitatively agrees with observations. 
    
\end{enumerate}



\section*{Acknowledgments}
We thank an anonymous referee for many constructive comments, which greatly improved the quality and clarity of our paper. 
Moreover, we thank Emma Willett for many useful discussions.
FV, AM, JTM, and JM acknowledge support from the European Research Council Consolidator Grant funding scheme (project ASTEROCHRONOMETRY, G.A. n. 772293). CK acknowledges funding from the United Kingdom 
Science and Technology Facility Council (STFC) through grant ST/R000905/1.
This work used the DiRAC Data Centric system at Durham University, operated by the Institute for Computational Cosmology on behalf of the STFC DiRAC HPC Facility (www.dirac.ac.uk). This equipment was funded by a BIS National E-infrastructure capital grant ST/K00042X/1, STFC capital grant ST/K00087X/1, DiRAC Operations grant ST/K003267/1 and Durham University. DiRAC is part of the National E-Infrastructure.
This research has made use of University of Hertfordshire's high-performance computing facility. We finally thank Volker Springel for providing \textsc{Gadget-3}.

\end{document}